\documentclass[aps,pre,10pt,twocolumn,preprintnumbers,amsmath,showkeys,showpacs]{revtex4}
\usepackage{lfmath} 
\usepackage{times}
\usepackage{mathptmx}
\usepackage{amsmath}
\usepackage{amsfonts}

\begin{document}

\title{Dissipation scales and anomalous sinks in steady two-dimensional turbulence}
\author{Eleftherios Gkioulekas}
\email{gkioulekase@utpa.edu}
\affiliation{Department of Mathematics, The University of Texas-Pan American , Edinburg, Texas, United States}
\begin{abstract}
In previous papers I have argued that the \emph{fusion rules hypothesis}, which was originally introduced by L'vov and Procaccia in the context of the problem of three-dimensional turbulence, can be used to gain a deeper insight in understanding the enstrophy cascade and inverse energy cascade of two-dimensional turbulence. In the present paper we show that the fusion rules hypothesis, combined with \emph{non-perturbative locality}, itself a consequence of the fusion rules hypothesis, dictates the location of the boundary separating the inertial range from the dissipation range. In so doing, the hypothesis that there may be an anomalous enstrophy sink at small scales and an anomalous energy sink at large scales emerges as a consequence of the fusion rules hypothesis. More broadly, we illustrate the significance of viewing inertial ranges as multi-dimensional regions where the fully unfused generalized structure functions of the velocity field are self-similar, by considering, in this paper, the simplified projection of such regions in  a two-dimensional space, involving a small scale $r$ and a large scale $R$, which we call, in this paper, the $(r, R)$-plane. We see, for example, that  the logarithmic correction in the enstrophy cascade, under standard molecular dissipation, plays an essential role in inflating the inertial range in the $(r, R)$  plane to ensure the possibility of local interactions. We have also seen that increasingly higher orders of hyperdiffusion at large scales or hypodiffusion at small scales make the predicted sink anomalies more resilient to possible violations of the fusion rules hypothesis.
\end{abstract}
\pacs{42.68.Bz, 47.27.-i, 47.27.ek, 92.60.hk, 92.10.ak}
\keywords{two-dimensional turbulence, fusion rules,  enstrophy cascade, inverse energy cascade}
\preprint{Preprint:  Submitted to \emph{Phys. Rev. E}}

\maketitle

\section{Introduction}

A very surprising property of Navier-Stokes turbulence in three-dimensions is that it has an anomalous energy sink. This means that in the forced-dissipative case, with energy  injected by random forcing at a constant average rate $\gee_{\text{in}}$, the rate of energy dissipation $\gee$ caused by the viscous term $\nu \del^2 u_\ga$ of the Navier-Stokes equation will eventually be equalized with $\gee_{\text{in}}$, even under the limit $\nu \to 0^+$ where the viscosity $\nu$ vanishes. This is surprising because one might expect that in the limit $\nu \to 0^+$, the Navier-Stokes equation should lose its capability to dissipate energy. This does not occur because as we decrease the viscosity $\nu$, the energy cascade adjusts by moving the dissipation length scale $\gn$, also known as the Kolmogorov microscale, further out into larger wavenumbers (i.e. smaller length scales). This intensifies the Laplacian $\del^2 u_\ga$ thus compensating for the decrease in the viscosity $\nu$ at  the dissipation term $\nu \del^2 u_\ga$. In textbooks (e.g. in Ref. \cite{book:Frisch:1995}), the location of the dissipation scale  $\gn$ is consequently derived by first assuming the existence of an anomalous energy sink and the existence of an energy cascade, and then deducing where the dissipation scale must be placed to dissipate the injected amount of energy. Thus, we obtain the well-known estimate $\gn \sim (\nu^3/\gee)^{1/4}$. 

The problem here  is that whereas the existence of an anomalous energy sink  has been well-established by experiments and numerical simulations \cite{article:Sreenivasan:1984,article:Water:2002,article:Uno:2003}, there is no mathematical proof, directly from first principles. A rigorous argument should be able to  establish from first principles: (a) that an energy cascade exists with energy spectrum scaling between $k^{-3}$ and $k^{-1}$; (b) the location of the dissipation scale $\gn$ itself. Then, from  (a) and (b), the existence of an anomalous energy sink follows.  This is obviously not an easy task. In fact, showing (a) is essentially almost the same thing as solving the problem of turbulence itself from first principles! From the standpoint of physics, it is fairly obvious that the reason why there is an anomalous energy sink in three-dimensional turbulence is the robust presence of a local energy cascade.  The difficulty in translating this physical intuition into a mathematical argument perhaps originates from the fact that we do not really have good grasp of what the proper mathematical definition for a stable local cascade should be in general.

The corresponding problem  of  whether  two-dimensional turbulence has an anomalous enstrophy sink at small scales and an anomalous energy sink at large scales, with certain caveats to be discussed below,  remains  open too. Short of a first-principles mathematical argument, as we shall show in this paper, it is relatively easy to formulate a weaker argument that addresses a weaker claim of the form: ``two-dimensional or three-dimensional turbulence will have anomalous sinks if and only if the fusion rules hypothesis (defined in Section \ref{sec:SummaryPrior}) is satisfied''. We can then go a step further and argue that ``the fusion rules hypothesis is satisfied if and only if two-dimensional or three-dimensional turbulence has universal self-similar scaling.'' The relationship between the fusion rules hypothesis and universality has been discussed in a previous paper \cite{article:Procaccia:1996:3} by L'vov and Procaccia, in the context of three-dimensional turbulence, and some of their results were generalized and extended to two-dimensional turbulence in my previous paper \cite{article:Gkioulekas:2008:1}. 

Before getting into the details of the matter, it is important to note that a proper investigation of the two-dimensional turbulence anomalous sinks problem is complicated by the apparent lack of robustness of the two-dimensional turbulence cascades   for which there is not currently a widely-accepted explanation \cite{article:Borue:1994,article:Gurarie:2001,article:Gurarie:2001:1,article:Danilov:2003,article:Shepherd:2002,article:Bowman:2003,article:Bowman:2004,article:Tung:2006}. For example, it is well-known, from the work of Tran and Bowman \cite{article:Shepherd:2002,article:Bowman:2003,article:Bowman:2004}, that the enstrophy cascade can fail to develop in the absence of a sufficiently strong large-scale sink. The large-scale sink is needed to dissipate the excess injected energy that cannot be disposed of by the small-scale sink. In the absence of a sufficiently strong large-scale sink,  both energy and enstrophy will pile up in the vicinity of the forcing wavenumber where they will  eventually be dissipated when the pile-up becomes sufficiently high. Likewise, for the inverse energy cascade, as Danilov and Gurarie \cite{article:Gurarie:2001,article:Gurarie:2001:1,article:Danilov:2003} have shown, the inverse energy cascade can be ``buried'' under  coherent vortices that hoard most of the energy of the system, resulting in a $k^{-3}$ contribution to the energy spectrum that dominates the $k^{-5/3}$  contribution of the still-existing inverse energy cascade. In a recent paper \cite{article:Gkioulekas:2007}, I proposed that these coherent structures originate from a self-amplification of the sweeping interactions part of the Navier-Stokes non-linearity. So far as we know, the inverse energy cascade coexists with these coherent structures, and can be recovered by artificially eliminating them \cite{article:Borue:1994,article:Gurarie:2001:1,article:Fischer:2005}. A theoretical explanation for this recoverability of the inverse energy cascade was given in previous papers \cite{article:Tung:2005,article:Tung:2005:1,article:Tung:2006}.

Contrast this behavior with the case of the energy cascade of three-dimensional turbulence. While it is also possible for the energy cascade to fail to develop, under a forcing-dissipation configuration where there is insufficient separation between the forcing scale and the dissipation scale, this can always be rectified by either increasing the rate of energy injection, or by decreasing the viscosity. In two-dimensional turbulence, a far more delicate tuning between forcing and the two sinks is required to recover a steady-state inverse energy cascade or a direct enstrophy cascade.  We are thus faced with the additional challenge of accounting for the conditions that are required for the existence of the cascades, in addition to addressing the existence of anomalous sinks, when these conditions are satisfied.

Further confusion arises from the fact that  under the Tran-Bowman scenario  \cite{article:Shepherd:2002,article:Bowman:2003,article:Bowman:2004}, we can have a trivial type of an anomalous energy sink simply by having energy pile up at the forcing range until the dissipation terms become strong enough to balance the rate of energy injection. In fact, it can be shown  that, at steady state, for finite viscosity, the total energy always has an upper bound \cite{article:Eyink:1996}. It follows that   two-dimensional turbulence always has an anomalous energy sink, even if it is merely of this trivial type. There is no known proof that a similar upper bound exists for the total enstrophy. As Eyink \cite{article:Eyink:1996} noted, if one  hypothesizes the existence of an upper bound for the total enstrophy, one may then predict that the placement of the dissipation scales gives a non-trivial anomalous enstrophy sink. However, in the absense of a proof for that bound, the existence of an anomalous enstrophy sink remains a completely open question. From the above remarks, we see that when we discuss the existence of anomalous sinks in the context of two-dimensional turbulence, we should qualify that our main interest is in the existence of a non-trivial anomalous energy sink that dissipates the energy in a dissipation range, located far away from the forcing range, so that the formation of an inverse energy cascade can be facilitated.  We are, likewise,  interested in  the existence of a non-trivial anomalous enstrophy sink, again located far away from the forcing range, that can facilitate an enstrophy cascade. 

Currently, there is a lot of interest in understanding this question of anomalous sinks in terms of the singular solutions of the Euler equations \cite{article:Eyink:2008}. In this paper we will explore a different route through the Navier-Stokes equations based on a mathematical framework that consists of the generalized balance equations governing the generalized structure functions $F_n$  and the fusion rules hypothesis. This mathematical framework was originally introduced by L'vov and Procaccia in the context of studying the energy cascade of three-dimensional turbulence \cite{lect:Procaccia:1994,article:Procaccia:1995:1,article:Procaccia:1995:2,article:Procaccia:1996,article:Procaccia:1996:1,article:Procaccia:1996:2,article:Procaccia:1996:3,article:Procaccia:1997,article:Procaccia:1998,article:Procaccia:1998:1,article:Procaccia:1998:2,article:Procaccia:2000}. In  earlier papers \cite{article:Tung:2005,article:Tung:2005:1},   I proposed that the method underlying this approach could be generalized to tackle the open questions that plague two-dimensional turbulence, and I used this approach to investigate  the locality and stability of the two-dimensional inverse energy cascade and downscale enstrophy cascade \cite{article:Gkioulekas:2008:1}. This investigation is continued in the present paper. 

The structure of the overall argument, in broad strokes, runs as follows: The generalized structure functions $F_n$ are defined as products of velocity differences where each velocity difference is evaluated at two points in space distinct from any other point-pair associated with the rest of the velocity differences. The usual standard structure functions, on the other hand, use the same point-pair for every velocity difference.  The fusion rules govern the scaling of these generalized structure functions when some but not all of the velocity point separations approach each other, while still remaining within the inertial range. We begin with the hypothesis  that there is a region of scales in which the generalized structure functions  satisfy incremental homogeneity, incremental isotropy, and a reasonably weak hypothesis of self-similarity, and the requirement that these symmetries  be universal. We consider this set of assumptions as a generalized abstract definition of the concept of a ``universal cascade''.  The first step of the argument is to derive the fusion rules from the universality hypothesis. Then, we proceed using the fusion rules, with  input from the governing Navier-Stokes equations, to show that: (1) within both inertial ranges  the nonlinear interactions are local and therefore a ``universal cascade'' driven by the Navier-Stokes equations has to be a ``local cascade''; (2) the self-similar scaling is indeed stable with respect to perturbations on the stochastic forcing term for the inverse energy cascade, and marginally stable for the downscale enstrophy cascade; (3) the self-similar scaling is not perturbed by the dissipation terms for some region of scales, and the dissipation scales are positioned as is necessary to provide for anomalous sinks that can dissipate the injected energy and enstrophy at wavenumbers far away from the forcing range. The formal setup of the argument and parts (1) and (2) were developed in a previous paper \cite{article:Gkioulekas:2008:1}. Part (3) is the subject of the present paper. This entire argument  has been carefully summarized in the conclusion of the present paper.

With respect to part (3) of the argument, our agenda in this paper, briefly stated, is as follows: We will show that in both two-dimensional and three-dimensional turbulence the fusion rules hypothesis implies the existence of anomalous sinks.  We will also show that a failure of the fusion rules hypothesis implies a corresponding failure of the anomalous sink hypothesis, with the caveat that this failure can be ameliorated by increasing the order of the corresponding dissipation operators. This  equivalence relation between the two hypotheses is interesting because using the fusion rules hypothesis as a point of departure makes it also possible to investigate the locality and stability of the  corresponding cascades. Our main focus in the present paper will be  the case of idealized  two-dimensional Navier-Stokes with linear dissipation both at small and large scales. However, the same argument easily carries over to three-dimensional turbulence, and that will be discussed here briefly as well. We will also derive some intermediate results regarding the dissipation scales of two-dimensional turbulence that are relevant to the question of the cascade stability.

The main thrust of the argument is to derive from the fusion rules the location of the dissipation scales both at the downscale range and the upscale range. We then use that to conclude that the scaling dependence of the upscale energy dissipation  $\gee_{ir}$ on the large-scale dissipation viscosity coefficient $\gb$ is given by 
\begin{equation}
\gee_{ir} \sim   \gb^{1-(\gz_2+2 m)/(\gz_2-\xi_{2,1}+2m)}.
\end{equation}
Likewise the dependence of the downscale enstrophy dissipation rate $\gn_{uv}$ on the viscosity coefficient $\nu$ is given by 
\begin{equation}
 \gn_{uv} \sim \nu^{1-(\gz_2-2(\gk+1))/(\xi_{2,1}-2(\gk+1))} [\ln (\ell_0/\gl_{uv})]^{a_3-1}.
\end{equation}
Here $\ell_0$ is the forcing scale and $\gl_{uv}$ is  the enstrophy dissipation scale. Furthermore,  $a_3$ is a scaling exponent associated with the  logarithmic scaling of the third-order vorticity structure function, $\gz_2$ is the  scaling exponent associated with the second-order velocity generalized structure function,   $\xi_{2,1}$ is the fusion  scaling exponent of that same structure function, $\gk$ is the order of the dissipation operator at small scales, and $m$  the order of the dissipation operator at large scales. Under the fusion rules hypothesis we have  $\xi_{2,1}=\gz_2$ for the downscale range,  and $\xi_{2,1}=0$  for the upscale range. For these values, the dissipation rates $\gn_{uv}$ and $\gee_{ir}$ become independent of the viscosities $\nu$ and $\gb$. Thus, under the fusion rules hypothesis we have non-trivial anomalous sinks. Furthermore, under the Falkovich-Lebedev prediction that $a_3 = 1$  \cite{article:Lebedev:1994,article:Lebedev:1994:1}, the enstrophy dissipation rate $\gn_{uv}$ also becomes independent of the logarithmic factor $\ln (\ell_0/\gl_{uv})$. We also see that increasing $\gk$ and $m$ leads to asymptotic independence of $\gn_{uv}$ and $\gee_{ir}$ from $\nu$ and $\gb$ even when the fusion scaling exponents deviate from the prescribed values.

Aside from the anomalous sink problem, the argument presented in this paper also sheds further light onto the problem of cascade stability discussed in the preceding paragraphs. The main idea is to re-envision the inertial range as a multidimensional region of scales in which the corresponding generalized structure functions retain self-similar scaling. Consider, for example, the case of the downscale enstrophy cascade. To first approximation let us assume that for the generalized structure function $F_n$, all velocity differences are evaluated at the length scale $R$ except for one pair evaluated at a smaller scale $r \ll R$ with both $r$ and $R$ in the inertial range. The crossover from the inertial range to the dissipation range occurs when $r$ is made small enough to be approximately equal to an $R$-dependent dissipation scale $\ell_{uv}^{(n)} (R)$. Our argument shows that the function $\ell_{uv}^{(n)} (R)$ can be calculated from the fusion rules hypothesis. The standard dissipation scale $\gl_{uv}^{(n)}$   that delineates the crossover from the inertial range to the dissipation range when all velocity difference pair separations are of the same length scale $R$ while being reduced simultaneously, is then obtained from the equation $\ell_{uv}^{(n)} (\gl_{uv}^{(n)}) = \gl_{uv}^{(n)}$. To have  non-trivial anomalous sinks, the dissipation scale $\gl_{uv}^{(n)}$   must have the correct leading-order dependence on the Reynolds number corresponding to the downscale cascade. More than that, we will argue that cascade stability with respect to the dissipation terms requires $\ell_{uv}^{(n)} (R)$ to satisfy the admissibility condition $a\gl_{uv}^{(n)} >  \ell_{uv}^{(n)} (a\gl_{uv}^{(n)})$ for all $a$ with $1<a<\ell_0/\gl_{uv}^{(n)}$.

For the upscale inverse energy cascade we begin with a generalized structure function $F_n$ where all velocity differences are evaluated at length scale $r$, and we expand one velocity difference to the length scale $R$ with $R\gg r$. The crossover to the dissipation range occurs when $R = \ell_{ir}^{(n)}(r)$, thereby defining the dissipation scale function $\ell_{ir}^{(n)}(r)$. Similarly, we define the standard dissipation scale $\gl_{ir}^{(n)}$  as the solution to the equation $\ell_{ir}^{(n)} (\gl_{ir}^{(n)}) = \gl_{ir}^{(n)}$. Cascade stability with respect to the dissipation terms again requires an admissibility condition, and for the inverse energy cascade it reads: $a\gl_{ir}^{(n)} <  \ell_{ir}^{(n)} (a\gl_{ir}^{(n)})$ for all $a$ with $\ell_0/\gl_{ir}^{(n)} < a < 1$.

 Let us now summarize our main results on cascade stability. For the case of the downscale enstrophy cascade we distinguish between the case where the cascade has intermittency corrections (i.e. $\gz_n = n+\gd_n$ with $\gd_2>0$ and $\gd_n<0$ for all $n$ with $n>3$) and the case of no intermittency corrections with molecular dissipation (i.e. $\gz_n = n$ and $\gk=1$). The fundamental difference between the two cases is that in the first case the dissipation scale function $\ell_{uv}^{(n)} (R)$ has a power-law leading-order dependence on the Reynolds number, whereas in the second case, the dependence becomes exponential to leading order. It should be noted that the mathematical argument for the first case also applies when there is the combination of no intermittency corrections and hyperdiffusion at small scales (i.e. $\gz_n = n$ and $\gk>1$), where the leading-order dependence of $\ell_{uv}^{(n)} (R)$ on the Reynolds number still follows a power-law.  For the first scenario of intermittency corrections we have shown that cascade stability with respect to the dissipation terms requires that  $\gz_2 < 2\gk$. It follows that under molecular diffusion $\gk=1$, a downscale enstrophy cascade with intermittency corrections would not be stable. For the second scenario of no intermittency corrections with molecular dissipation (i.e. $\gz_n = n$ and $\gk=1$) we find that the downscale enstrophy cascade is stable with respect to the dissipation terms provided that  $\gl_{uv}^{(n)} \ll \ell_0/\sqrt{e}$. For hyperdiffusion (i.e. $\gz_n = n$ and $\gk>1$) this minor constraint is removed.

As for the inverse energy cascade, we find that cascade stability with respect to the dissipation terms requires that the scaling exponents $\gz_n$  of the generalized structure functions $F_n$ must satisfy the inequality  $\gz_{n+1}-\gz_n < 2m+1$. Given that we do not know whether the inverse energy cascade has intermittency corrections, the status of this inequality is uncertain. It should be noted that in the above, cascade stability is understood in the strong sense of requiring \emph{all} generalized structure functions $F_n$ to have an inertial range with universal scaling. These results add to our previous results on cascade stability with respect to the forcing terms \cite{article:Gkioulekas:2008:1}.

 Combined with the results of my previous paper \cite{article:Gkioulekas:2008:1}, we are beginning to see a big picture in which the fusion rules hypothesis operates as a unifying nexus that can subsume 3 distinct assumptions that everyone makes about turbulence cascades: locality, stability, and the existence of anomalous sinks. This unification provides us with two opportunities. First, it is possible to investigate the validity of the fusion rules hypothesis with numerical simulations, which are easier to do in two dimensions. Second, and more important, the validity of the fusion rules can be investigated further on theoretical grounds, by generalizing similar arguments proffered for the problem of three-dimensional turbulence \cite{lect:Procaccia:1994,article:Procaccia:1995:1,article:Procaccia:1995:2,article:Procaccia:1996} and the passive scalar problem \cite{article:Procaccia:2000:1}. It is worth noting that experiments have corroborated the fusion rules both for three-dimensional turbulence \cite{article:Sreenivasan:1997,article:Procaccia:1998:3,article:Toschi:1998,article:Toschi:1999,article:Reeh:2000,article:Tabar:2000}  and for the passive scalar problem \cite{article:Procaccia:1996:5,article:Procaccia:1996:6}.

This paper is organized as follows. Section II reviews the framework of the generalized balance equations and the fusion rules hypothesis. In Section III we show how the law governing the location of dissipation scales follows as a consequence of the fusion rules hypothesis. Underlying this argument is a geometrical conception of either inertial range as a two-dimensional region separated by the corresponding dissipative region, by a curve whose shape is deduced from the fusion rules hypothesis and locality (itself a consequence of the fusion rules hypothesis). The curvature of that line provides us with an admissibility condition and also leads to the standard dissipation scale which is relevant to the anomalous sink question. In Section IV we apply the method of Section III for the case of the enstrophy cascade, and in Section V we consider the case of the inverse energy cascade. Having derived the laws governing the standard dissipation scales from the fusion rules for both cascades, in Section VI we turn to the question of anomalous sinks. Conclusions and discussion, summarizing the logical structure of the argument as a whole, are given in Section VII. Technical matters are taken up in the appendices.

\section{Summary of prior results}
\label{sec:SummaryPrior}


The generalized balance equations were originally derived by L'vov and Procaccia \cite{article:Procaccia:1996:3}  and, combined with the fusion rules hypothesis, they are the foundation of the non-perturbative L'vov-Procaccia theory of three-dimensional turbulence \cite{article:Procaccia:1996:1,article:Procaccia:1996:2,article:Procaccia:1996:3,article:Procaccia:1998,article:Procaccia:1998:1,article:Procaccia:1998:2}, and also the foundation for a corresponding investigation of the cascades of two-dimensional turbulence \cite{article:Tung:2005,article:Tung:2005:1,article:Gkioulekas:2008:1}. We begin, in this section, by summarizing the main results from our previous paper \cite{article:Gkioulekas:2008:1} on the generalized balance equations and the fusion rules hypothesis.

\subsection{The balance equations}

We begin with the Navier-Stokes equation for the velocity field in two dimensions:
\begin{equation}
\pderiv{u_{\ga}}{t} + \cP_{\ga\gb}\pdc (u_{\gb} u_{\gc}) = \cD u_{\ga} + \cP_{\ga\gb} f_{\gb}.
\end{equation}
Here, $u_\ga$ is the Eulerian velocity field, $\pd_\ga$ the partial spatial derivative in the $\ga$-direction, $\cP_{\ga\gb}$ is the projection operator $\cP_{\ga\gb} \equiv \gd_{\ga\gb} - \pda\pdb\ilapl$, $f_\ga$ is the forcing term, and $\cD$ is the dissipation operator  given by
\begin{equation}
\cD  \equiv (-1)^{\gk+1} \nu\del^{2\gk}   + (-1)^{m+1}\gb \del^{-2m}.
\end{equation}
Here the integers $\gk$ and $m$ describe the order of the dissipation mechanisms, and  the numerical coefficients $\nu$ and $\gb$ are the corresponding viscosities.   The first term in $\cD$ is the small-scale sink, and the second term is the large-scale sink. The case $\gk = 1$ corresponds to  standard molecular viscosity, and the case $m=0$ corresponds to Ekman damping.  

From the Navier-Stokes equations  we  derive the exact statistical theory of velocity differences. Let  $w_{\alpha}(\bfx ,\bfxp ,t)$  be the Eulerian velocity difference, defined as:  
\begin{equation}
w_{\alpha}(\bfx ,\bfxp ,t) = u_{\alpha}(\bfx, t) - u_{\alpha}(\bfxp ,t).
\end{equation}
To write equations  concisely, we introduce the following notation to represent aggregates of position vectors 
\begin{align}
\bfX &= (\bfx , \bfxp), \\ 
\{\bfX \}_n &= \{\bfX_{1} , \bfX_{2} , \ldots , \bfX_{n}\},\\ 
\{\bfX\}_n^k &= \{\bfX_{1} , \ldots , \bfX_{k-1} , \bfX_{k+1} , \ldots , \bfX_{n}\}.
\end{align}
Below, the notation $\|\{\bfX\}_n\| \sim R$ means that all point to point distances in the geometry of velocity differences $\{\bfX\}_n$ have the same order of magnitude $R$. Similarly, the notation $\|\{\bfX\}_n\| \ll \|\{\bfY\}_n\|$ means that all the point to point distances in  $\{\bfY\}_n$ are much larger than all the point to point distances in $\{\bfX\}_n$. 

The eulerian one-time fully unfused correlation tensors are formed by multiplying $n$ velocity differences $w_{\alpha}(\bfx ,\bfxp ,t)$ evaluated at $2n$ distinct points $\{\bfX \}_n$:
\begin{equation}
F_n(\{\bfX \}_n, t) = \left\langle \left [  \prod_{k=1}^n w_{\ga_k} (\bfX_{k} , t) \right] \right\rangle.   
\end{equation} 
Differentiating  $F_n$ with respect to $t$ and applying the Navier-Stokes equations yields  equations of the form
\begin{equation}
\pderiv{F_n}{t} + D_n =  \nu J_n +\gb H_n +Q_n, 
\end{equation}
where $D_n$ is the combined contribution of the pressure and the nonlinear term, $Q_n$ is the contribution of the forcing term, $J_n$ accounts for diffusion or hyperdiffusion, and $H_n$ accounts for large-scale dissipation. We call these equations the \emph{generalized balance equations}.  The dissipation terms are given by
\begin{equation}
\begin{split}
H_n (\{\bfX \}_n, t) &= \sum_{k=1}^n  (\nabla^{-2m}_{\bfx_k} + \nabla^{-2m}_{\bfxp_k})F_n (\{\bfX \}_n, t), \\
J_n (\{\bfX \}_n, t) &= \sum_{k=1}^n  (\nabla^{2\gk}_{\bfx_k} + \nabla^{2\gk}_{\bfxp_k}) F_n (\{\bfX \}_n, t),
\end{split}
\end{equation}
where $\nabla^{2\gk}_{\bfx_k}$ is a hyperlaplacian operator that differentiates with respect to $\bfx_k$ and similarly with $\nabla^{2\gk}_{\bfxp_k}$, $\nabla^{-2m}_{\bfx_k}$ and $\nabla^{-2m}_{\bfxp_k}$.  The forcing contribution is given by
\begin{equation}
\begin{split}
Q_n (\{\bfX\}_n, t) &= \sum_{k=1}^n Q_{kn}(\{\bfX\}^k_n , \bfX_k, t),\\ 
Q_{kn}(\{\bfX\}_{n-1} ,\bfY, t) &= \left\langle \left [  \prod_{k=1}^{n-1} w_{\ga_k} (\bfX_k , t) \right] \gf_\gb (\bfY, t) \right\rangle ,
\end{split}
\end{equation}
where
\begin{equation}
\gf_\ga (\bfX, t)= f_{\alpha}(\bfx, t) - f_{\alpha}(\bfxp ,t).
\end{equation}
The non-linear term $D_n$ can be rewritten as
\begin{equation}
D_n (\{\bfX\}_n, t) = \sum_{k=1}^n D_{kn}(\{\bfX\}_n , t) + I_n (\{\bfX\}_n, t),
\end{equation}
with $I_n$ representing the \emph{sweeping interactions} given by
\begin{multline}
I_n (\{\bfX\}_n, t)  \\ =\sum_{k=1}^n (\partial_{\gb, \bfx_{k}} +\partial_{\gb, \bfxp_{k}} )\left\langle \cU_\gb (\{\bfX\}_n, t) \left [  \prod_{k=1}^n w_{\ga_k} (\bfX_{k} , t) \right] \right\rangle, 
\end{multline}
where $\cU_\gb (\{\bfX\}_n, t)$ is the mean velocity field associated with the configuration $\{\bfX \}_n$  defined as
\begin{equation}
\cU_\gb (\{\bfX\}_n, t) = \frac{1}{2n}\sum_{k=1}^n \left (u_{\alpha}(\bfx_{k}, t)+  u_{\alpha}(\bfxp_{k} ,t) \right ).
\end{equation}
The term $D_{kn}$ represents the \emph{local non-linear interactions} given, in general form, via a linear integrodifferential operator $\cO$, as follows:
\begin{multline}
D_{kn} (\{\bfX\}_n, t)  \\ = \iint \df{\bfY_1} \df{\bfY_2} \; \cO (\bfX_k ,\bfY_1 ,\bfY_2) \;F_{n+1} (\{\bfX\}^k_n, ,\bfY_1 ,\bfY_2,t).
\end{multline}
The detailed form of the local nonlinear term $D_{kn}$ is tedious and was given in our previous paper \cite{article:Gkioulekas:2008:1}. 

\subsection{Self-similar scaling and the fusion rules}

For a stationary problem, the generalized balance equations can be rewritten as 
\begin{equation} 
\cO_n F_{n+1} + I_n =  \nu J_n +\gb H_n + Q_n. 
\end{equation}

As we have explained in our previous paper \cite{article:Gkioulekas:2008:1}, to considerable detail, the downscale or upscale inertial range of two-dimensional turbulence can be identified with a region $\cJ_n \subseteq \bbR^{2n}$ in which for $\{\bfX\}_n\in \cJ_n$, the contributions of the nonlinear term $\cO_n F_{n+1}$ dominate all the other terms of the balance equations. The  region $\cJ_n$  is circumscribed on one side by the forcing term $Q_n$ and on the other side by the corresponding dissipation terms $\nu J_n$ and $\gb H_n$. Meanwhile, we assume, for the time being without further proof, that the sweeping term $I_n$ also remains negligible with respect to $\cO_n F_{n+1}$ in the region $\cJ_n$. A preliminary study of the sweeping term was given in another  paper \cite{article:Gkioulekas:2007}. 

Within the region $\cJ_n$ we expect that $F_n$ will be self-similar according to the following scaling law:
\begin{equation}
F_n(\gl\{\bfX\}_n, t) =\gl^{\gz_n} F_n(\{\bfX\}_n, t).
\end{equation}
This scaling  law is expected to hold when $\{\bfX\}_n\in \cJ_n$, and the scaling exponents $\gz_n$ characterize the corresponding inertial range, and will obviously differ between the upscale range  and the downscale range. Within the same region $\cJ_n$ it is expected that  $F_n$ will be incrementally homogeneous and incrementally isotropic. 

We also introduced \cite{article:Gkioulekas:2008:1} the hypothesis that both cascades of two-dimensional turbulence, when they exist, will satisfy another group of self-similarity laws called the \emph{fusion rules}. We would now  like to  briefly summarize the content of these rules.  Consider a geometry of velocity differences $\{\bfx\}_n$ such that all point to point distances have order of magnitude $1$, and define 
\begin{equation}
F_n^{(p)} (r,R) = F_n(r\{\bfx_k\}_{k=1}^{p},R\{\bfx_k\}_{k=p+1}^{n}).
\end{equation}
The function $F_n^{(p)} (r,R)$ reflects the case where $p$ velocity differences have separations with order of magnitude $r$, and $n-p$ velocity differences have separations with order of magnitude $R$. The case of interest is when the evaluation $(r\{\bfx_k\}_{k=1}^{p},R\{\bfx_k\}_{k=p+1}^{n})$ is within the inertial range  $\cJ_n$ and $r\ll R$.  The fusion rules  give the scaling properties of $F_n^{(p)}$. 

We distinguish between two cases. For a direct cascade, such as the energy cascade of three-dimensional turbulence and the enstrophy cascade of two-dimensional turbulence, the fusion rules are given by 
\begin{equation}
F_n^{(p)} (\gl_1 r,\gl_2 R) = \gl_1^{\xi_{n,p}}\gl_2^{\gz_n-\xi_{n,p}} F_n^{(p)} (r,R),
\end{equation}
for $2\le p\le n-2$ with $\xi_{n,p}=\gz_p$. In my previous paper \cite{article:Gkioulekas:2008:1}, I also introduced the notion of a regular violation of the fusion rules, if the fusion exponents $\xi_{n,p}$ satisfy the inequality $0\leq \xi_{n,p}\leq \gz_n$, which is necessary and sufficient to ensure that $\gl_1$ and $\gl_2$ are not governed by negative exponents, since, by definition, $F_n$ should vanish in the limits $\gl_1\to 0$ or  $\gl_2\to 0$. For the cases $p=1$ and $p=n-1$ the fusion exponents are given by  $\xi_{n,1}=\gz_2$ and $\xi_{n,n-1}=\gz_n$. For the case $p=1$ it is assumed that one of the end points of the small velocity difference is attached to one of the big velocity differences, that is $\bfx_1 = \bfx_2$. This assumption holds when working with partially fused correlations, where all point-pairs share a point in common.  It also holds when integrating $F_2$ to compute the energy spectrum.

For the inverse energy cascade of two-dimensional turbulence, as was shown in my previous paper \cite{article:Gkioulekas:2008:1}, the fusion exponents are given instead by $\xi_{n,p}=\gz_n-\gz_{n-p}$ for $2\le p\le n-2$. For the cases $p=1$ and $p=n-1$ the fusion exponents are given by $\xi_{n,n-1}=\gz_n-\gz_2$ and $\xi_{n,1}=0$. Again, the case $p=n-1$ requires the assumption that the large velocity difference is attached to one of the small velocity differences, that is $\bfx_n = \bfx_{n-1}$.  Note that, compared to the direct cascade, the roles of small and large velocity differences are switched.

The fusion rules can be shown to be a consequence of the universality hypothesis, which  we have introduced previously \cite{article:Gkioulekas:2008:1} in a generalized form that is applicable to both the upscale and the downscale ranges of two-dimensional turbulence. According to the universality hypothesis, the conditional generalized structure functions $\Phi_n$, given by
\begin{multline}
\gF_n(\{\bfX\}_n,\{\bfY\}_m,\{\bfw_k\}_{k=1}^{m}, t) \\ = \left\langle  \left. \left [  \prod_{k=1}^n w_{\ga_k} (\bfX_{k} , t) \right]  \right|  \bfw(\bfY_k,t)=\bfw_k, \;\forall k\in \{1,\ldots,m\})\right\rangle,
\end{multline}
will honor the same symmetries, with respect to the point-pairs  $\{\bfX\}_n$, as the unconditional correlations $F_n$, in the asymptotic limit where  $\|\{\bfY\}_m\|$ is situated between $\|\{\bfX\}_n\|$ and the forcing scale $\ell_0$ (i.e. $\ell_0 \gg \|\{\bfY\}_m \| \gg \|\{\bfX\}_n \|$ for a downscale cascade; $\ell_0 \ll \|\{\bfY\}_m \| \ll \|\{\bfX\}_n \|$ for an upscale cascade). These symmetries are: incremental homogeneity, incremental isotropy, and self-similarity with the same scaling exponents $\gz_n$. The ensemble here is constrained by a control condition imposed on the velocity differences $\bfw(\bfY_k,t)$. It can be shown that the values of the fusion rules scaling exponents $\xi_{n,p}$ can be derived as a consequence of this hypothesis of universality for both inertial ranges \cite{article:Procaccia:1996:1,article:Procaccia:1996:3,article:Gkioulekas:2008:1}. A better formulation of the universality hypothesis would use a control on the forcing term $\gf_\ga (\bfY, t)$ instead of a control on the velocity differences $\bfw(\bfY_k,t)$. Unfortunately, this weakens the hypothesis so much that obtaining the fusion rules exponents $\xi_{n,p}$ becomes a very challenging problem. We discuss this important matter in the conclusion of this paper.

From the fusion rules we can  show that the integrals in the nonlinear interactions term  $\cO_n F_{n+1}$ are local, which implies that the scaling exponent of $\cO_n F_{n+1}$ is $\gz_{n+1}-1$.  If the fusion rules fail, then as long as the violation is regular (i.e. $0<\xi_{n,p}<\gz_n$), UV localiy is still maintained. Let us assume that  $\xi_{n,p} = \gz_p+\gD \xi_{n,p}$ in the downscale range and $\xi_{n,p} = \gz_n-\gz_{n-p}+\gD \xi_{n,p}$ in the upscale range, with $\gD \xi_{n,p}$ the corresponding perturbation. Then, IR  locality  is also  maintained if the deviations in the fusion exponents  satisfy $\gD \xi_{n+1,2} + \gD \xi_{n+1,n-1} \geq 0$ (downscale range) or $\gD \xi_{n+1,2} + \gD \xi_{n+1,n-1} \leq 0$ (upscale range), for $n>1$. If both UV and IR locality are maintained, then the scaling exponent of $\cO_n F_{n+1}$ is still $\gz_{n+1}-1$.

Knowing the scaling of the nonlinear term, it becomes possible to compare it against the other terms of the balance equations. As we have shown in our previous paper  \cite{article:Gkioulekas:2008:1}, if we assume that the forcing term is random-gaussian, then we may calculate the scaling exponent of $Q_n$ and compare it against $\cO_n F_{n+1}$. Thus, one  can effectively circumscribe the boundary between the region $\cJ_n$ and the forcing range. In the present paper we will show how to compare the nonlinear term against the dissipation terms $\nu J_n$ and $\gb H_n$. In doing so, we  determine the position of the corresponding dissipation scales and may then show that they are positioned consistently with the anomalous sinks hypothesis. Thus, we will show that the anomalous sinks hypothesis can be derived as a consequence of the fusion rules hypothesis, which in turn follows from  the proposed universality hypothesis.

\section{Dissipation scales from the fusion rules}

The obvious problem with estimating the position of the dissipation scales by dimensional analysis is that by doing so we presuppose the validity of a correspoding anomalous sink hypothesis. The argument that we will use instead is similar to the argument of L'vov-Procaccia \cite{article:Procaccia:1996:2} for the case of the energy cascade of three-dimensional turbulence.  We now review this method, show how it can be extended for the inverse energy cascade, and present an additional consideration that is very relevant in two-dimensional turbulence. 

Consider the case where in the correlation $F_n^{(1)} (r,R)$ between $n$ velocity differences, one of them is evaluated at separation $r$ and all others at $R$ such that $r\ll R\ll \ell_0$. Here, we assume that both $r$ and $R$ are still in the inertial range.  For sufficiently small $r$, universal scaling fails as one leaves the inertial range and enters  the dissipation range.  The scale $\ell_{uv}^{(n)}$ where this cross-over occurs is dependent on $R$. Thus, one may define a dissipative length scale function $r=\ell_{uv}^{(n)}(R)$ whose graph traces out the dissipative boundary of the enstrophy inertial range in the $(r,R)$ plane. Similarly, in the case where $\ell_0 \ll r \ll R$, one may define a large-scale dissipation length scale function $R=\ell_{ir}^{(n)}(r)$ that marks the crossover point where the function $F_n^{(n-1)} (r,R)$ enters the large-scale dissipation range when $R$ exceeds $\ell_{ir}^{(n)}(r)$.  The function $\ell_{ir}^{(n)}(r)$ sketches out the shape of the inverse energy cascade in the $(r,R)$ plane.  It should be remembered that a complete representation of the extent of one of the inertial ranges requires at least an $2n$ dimensional region $\cJ_n \subseteq \bbR^{2n}$ for the correlation $F_n$.  Nevertheless, this two-dimensional representation is sufficient for our purposes, and certainly a step forward from the usual one-dimensional representation.

In the enstrophy inertial range, when $r$ is still in the inertial range, the function $F_n^{(1)} (r,R)$ evaluates according to the $p=1$ fusion rule, and it is given by 
\begin{equation}
F_n^{(1)} (r,R)\sim \frac{F_2 (r)F_n (R)}{F_2 (R)}.
\end{equation}
Since we are ultimately interested in the dissipation length scales of structure functions, we may safely assume that one end point of the small velocity difference coincides with one of the end points of another velocity difference. The validity of the $p=1$ fusion rule, used above, is contigent on making this assumption. When $r$ enters the dissipation range, the dominant balance in the balance equations is 
\begin{equation}
\nu\del^{2\gk}_{\bfx_1} F_n^{(1)} (r,R) \sim \frac{F_{n+1}(R)}{R}, 
\end{equation}
and solving for $F_n^{(1)} (r,R)$ yields
\begin{equation}
F_n^{(1)} (r,R)\sim \frac{r^{2\gk}F_{n+1}(R)}{\nu R}.
\end{equation}
The dissipation length scale function is determined by matching the two asymptotic expressions for $F_n^{(1)} (r,R)$:
\begin{equation}
\frac{F_2 (\ell_{uv}^{(n)}(R))F_n (R)}{F_2 (R)}\sim \frac{(\ell_{uv}^{(n)}(R))^{2\gk}F_{n+1}(R)}{\nu R},
\end{equation}
and solving for $\ell_{uv}^{(n)}(R)$. The next step is to determine the length scale  $\gl_{uv}^{(n)}$ where   a cross-over to the dissipation range occurs when all  velocity differences are shrinked simultaneously. We estimate the location of this cross-over scale by solving the equation $\ell_{uv}^{(n)}(\gl_{uv}^{(n)})=\gl_{uv}^{(n)}$ with respect to $\gl_{uv}^{(n)}$. If the enstrophy cascade forms successfully, then the dissipation length scale of the energy spectrum will be approximately located at $\gl_{uv}^{(2)}$, thus $\gl_{uv}^{(2)}$ is effectively an estimate of the Kolmogorov microscale. Similarly, the dissipation scales of the standard structure functions $S_n$ will be approximately at $\gl_{uv}^{(n)}$.  In this sense, we say that $\gl_{uv}^{(n)}$ are the \emph{standard} dissipation length scales. 

Although the standard dissipation scales are the ones that we are ultimately interested in, the shape of the function $\ell_{uv}^{(n)}(R)$ is also significant in the following sense: Suppose that all the velocity differences have been shrinked down to the scale $\gl_{uv}^{(n)}$, and we begin to stretch simultaneously $n-1$ velocity differences by a factor $a>1$ to length $a\gl_{uv}^{(n)}$ while adjusting the remaining velocity difference to length $\ell_{uv}^{(n)}(a\gl_{uv}^{(n)})$ such that we remain on the boundary between the inertial range and the dissipation range.  Then, the shape of the function $\ell_{uv}^{(n)}$ has to be such that we may remain on the boundary  without being forced to increase the separation of the small velocity difference more than we have increased the other $n-1$ velocity differences.  In other words, we propose that in a valid inertial range it is necessary that the following admissibility condition be satisfied:
\begin{equation}
a\gl_{uv}^{(n)} > \ell_{uv}^{(n)}(a\gl_{uv}^{(n)}) \; ,\forall a \in (1,\ell_0/\gl_{uv}^{(n)}).
\end{equation}
When this condition fails, we can have the inconsistent situation of exiting the inertial range and entering the dissipation range if we increase all velocity differences separations  simultaneously by the same factor! This condition is obviously satisfied when $\ell_{uv}^{(n)}$ is a decreasing function.  However, it can also be satisfied when $\ell_{uv}^{(n)}$ is increasing, as long as the curve $r=\ell_{uv}^{(n)}(R)$ remains underneath the line $r=R$. It will be shown later that this condition fails for the enstrophy range in certain cases. 

The same technique can be applied to the inverse energy cascade inertial range.  The evaluations of the function $F_n^{(n-1)} (r,R)$ in the inertial range and the dissipation range are given by 
\begin{equation}\begin{split}
F_n^{(n-1)} (r,R)&\sim \frac{F_2 (R)F_n (r)}{F_2 (r)},\\
F_n^{(n-1)} (r,R)&\sim \frac{R^{-2m}F_{n+1}(r)}{\gb r},
\end{split}
\end{equation}
and the dissipation scale function $\ell_{ir}^{(n)}(r)$ is found by matching the two equations of $F_n^{(n-1)} (r,R)$  and solving  the equation
\begin{equation}
\frac{F_2 (\ell_{ir}^{(n)}(r))F_n (r)}{F_2 (r)}\sim \frac{(\ell_{ir}^{(n)}(r))^{-2m}F_{n+1}(r)}{\gb r}.
\end{equation}
Then we obtain the standard dissipation scale $\gl_{ir}^{(n)}$  by solving the equation $\ell_{ir}^{(n)}(\gl_{ir}^{(n)})=\gl_{ir}^{(n)}$ and we demand that the shape of the function $\ell_{ir}^{(n)}$ should satisfy the condition
\begin{equation}
a\gl_{ir}^{(n)} < \ell_{ir}^{(n)}(a\gl_{ir}^{(n)}) \; ,\forall a \in (\ell_0/\gl_{ir}^{(n)},1).
\end{equation}

We would like to stress again the significance of the form of the functions $\ell_{uv}^{(n)}$ and $\ell_{ir}^{(n)}$ from another point of view.  Consider for example the case of the enstrophy cascade.  Suppose that we plot the lines $r=R$ and $r=\ell_{uv}^{(n)}(R)$ on the $(r,R)$ plane.  These lines intersect at the standard dissipation scale $\gl_{uv}^{(n)}$ and they may or may not intersect elsewhere.  The region between the line $r=\ell_{uv}^{(n)}(R)$ and  the line $r=R$ can be thought of as a two-dimensional representation of the region where dissipation is negligible. Every point within that region represents a set of velocity differences configurations for which the correlations maintain self-similar universal scaling.   To actually have an inertial range, the curve $r=\ell_{uv}^{(n)}(R)$ has to be underneath the line $r=R$ from the point of intersection at the standard dissipation scale $\gl_{uv}^{(n)}$ at least up until the forcing scale $\ell_0$; the region between them needs to be reasonably inflated to allow local interactions to take place.  A similar argument is applicable for the case of the inverse energy cascade. We see therefore that these admissibility conditions are essentially cascade stability conditions with respect to the dissipation terms. As such, they complement the cascade stability conditions with respect to the forcing terms that we discussed in our previous paper \cite{article:Gkioulekas:2008:1}.

\section{The case of the enstrophy cascade}

We now employ the method described in the previous section  to study the geometry of the enstrophy range in the $(r,R)$ plane.  In this section, we let $\gn_{uv}$  represent  the amount of the enstrophy that flows downscale.  The value of $\gn_{uv}$ is the response of the system to the combined forcing and large-scale dissipation that drive the enstrophy cascade.  A Reynolds number can be defined for the enstrophy range as
\begin{equation}
\cR^{(\gn)}_{uv} =\frac{\gn_{uv}^{1/3}\ell_0^{2\gk}}{\nu}. 
\end{equation}
We will consider two possibilities: an enstrophy inertial range with intermittency corrections and a logarithmic enstrophy range without intermittency corrections.  Although the possibility of intermittency corrections was recently ruled out by Eyink \cite{article:Eyink:2001}, it is still instructive to consider the possibility on a hypothetical basis. The reason for doing so is because we will show that  in the case $\gk=1$,  corresponding to  standard molecular diffusion, the geometry of the boundary between the inertial range and the dissipation range offers sufficient grounds to reject the intermittency scenario! 

Because the argument below is somewhat technical, we will begin by first summarizing our main results. For the first case of an enstrophy cascade with intermittency corrections (i.e. $\gz_n = n+\gd_n$ with $\gd_2>0$ and $\gd_n<0$ for all $n$ with $n>3$), the dissipation scale function $\ell_{uv} (R)$ is given by 
\begin{equation}
\frac{\ell_{uv}(R)}{\ell_0} = \left [ \frac{\cR^{(\gn)}_{uv}}{\cR^{(\gn)}_{2,uv}} \left( \frac{R}{\ell_0}\right)^2 \right]^{1/(\gz_2-2\gk)}, 
\end{equation}
where $\cR^{(\gn)}_{2,uv}$ is the critical Reynolds number. Note that this result also applies when $\gz_2=2$ (no intermittency corrections) and $\gk>1$ (hyperdiffusion). Solving the equation $\ell_{uv} (\gl_{uv}) = \gl_{uv}$ gives the standard dissipation scale $\gl_{uv}$, which reads 
\begin{equation}
\left(\frac{\gl_{uv}}{\ell_0}\right)= \left [ \frac{\cR^{(\gn)}_{uv}}{\cR^{(\gn)}_{2,uv}}  \right]^{1/(\gz_2-2(\gk+1))}. 
\end{equation}
The admissibility condition $a\gl_{uv} >  \ell_{uv} (a\gl_{uv})$ requires that $\gz_2<2\gk$. For the higher-order generalized structure functions, the corresponding dissipation scale function $\ell_{uv}^{(n)} (R)$ is given by 
\begin{equation}
\frac{\ell_{uv}^{(n)}(R)}{\ell_0} = \left( \frac{R}{\ell_0}\right)^{x_n} \left [ \frac{\cR^{(\gn)}_{uv}}{\cR^{(\gn)}_{n,uv}}  \right]^{1/(\gz_2-2\gk)},
\end{equation}
with the scaling exponents $x_n$ given by 
\begin{equation}
x_n = \frac{\gz_{n+1}-\gz_n+\gz_2 -1}{\gz_2-2\gk}. 
\end{equation}
The standard dissipation scale $\gl_{uv}^{(n)}$ then reads
\begin{equation}
\frac{\gl_{uv}^{(n)}}{\ell_0} = \left [\frac{\cR^{(\gn)}_{uv}}{\cR^{(\gn)}_{n,uv}}\right]^{1/[1-2\gk-(\gz_{n+1}-\gz_n)]}.
\end{equation}
The admissibility condition for $F_n$ gives again the  condition $\gz_2<2\gk$. 

A distinct argument is needed for the case $\gz_2=2$  and $\gk=1$ because then the dissipation scale function $\ell_{uv} (R)$ changes from a power-law dependence on the Reynolds number $\cR^{(\gn)}_{uv}$  to exponential dependence. Using the evaluation $F_n (R) \sim (\gn_{uv}^{1/3}R)^n [\ln (\ell_0/R)]^{a_n}$,  we find that the dissipation scale function $\ell_{uv}^{(n)} (R)$ is given by 
\begin{equation}
\frac{\ell_{uv}^{(n)}(R)}{\ell_0}=
\exp\left [ - \fracp{\cR^{(\gn)}_{uv}}{\cR^{(\gn)}_{2,uv}}^{1/a_2}
\fracp{R}{\ell_0}^{2/a_2}
\left [\ln\fracp{\ell_0}{R}\right]^{b_n}\right ],
\end{equation}
with $b_n$ given by 
\begin{equation}
b_n = \frac{a_{n+1}-a_n+a_2}{a_2}.
\end{equation}
The standard dissipation scale $\gl_{uv}^{(n)}$  is found by solving the transcendental equation 
\begin{equation}
\fracp{\ell_0}{\gl_{uv}^{(n)}}^{2/a_2}
\left [\ln \fracp{\ell_0}{\gl_{uv}^{(n)}}\right]^{1-b_n} = \fracp{\cR^{(\gn)}_{uv}}{\cR^{(\gn)}_{2,uv}}^{1/a_2}.
\end{equation}
The admissibility condition is satisfied given sufficient separation between  $\gl_{uv}^{(n)}$ and $\ell_0$ 

We now proceed with the detailed derivation and discussion of the above results.


 \subsection{Enstrophy range with intermittency corrections} 

As a point of departure, we assume the evaluation given by the $1/8$ law that $F_3 (r)\sim (\gn_{uv} r^3). $ A derivation of the $1/8$ law is given by Lindborg \cite{article:Lindborg:1999}, Bernard \cite{article:Bernard:1999}, and Davidson \cite{article:Davidson:2008}.  
For the other correlations we will allow the general form
\begin{equation}
F_n (r) \sim (r/\ell_0)^{\gz_n -n}(\gn_{uv}^{1/3}r)^n,
\end{equation}
where the scaling exponents $\gz_n$ satisfy $\gz_n = n+\gd_n$ with $\gd_2>0$ and $\gd_n<0$ for all $n$ with $n>3$. It follows from the $p=1$ fusion rule that the evaluation of the correlation $F_2^{(1)}(r,R)$, when $r$ is in the inertial range and the dissipation range correspondingly, reads
 \begin{equation}
 \begin{split}
 F_2^{(1)}(r,R) &\sim F_2 (r) \sim  (r/\ell_0)^{\gz_2 -2}(\gn_{uv}^{1/3}r)^2,\\
 F_2^{(1)}(r,R) &\sim \frac{r^{2\gk}F_{3}(R)}{\nu R} \sim \frac{r^{2\gk} R^2}{\nu}.
 \end{split}
 \end{equation}
These evaluations coincide at the boundary between the inertial range and the dissipation range.  For a fixed $R$, the corresponding dissipation length scale $\ell_{uv}(R)$ is found from the following matching condition
\begin{equation}
(\ell_{uv}(R)/\ell_0)^{\gz_2 -2}(\gn_{uv}^{1/3}\ell_{uv}(R))^2 \sim \frac{[\ell_{uv}(R)]^{2\gk} R^2}{\nu},
\end{equation}
and solving for $\ell_{uv}(R)$ we obtain
\begin{equation}
\frac{\ell_{uv}(R)}{\ell_0} \sim \left [ \frac{\gn_{uv}^{1/3}\ell_0^{2\gk}}{\nu} \left( \frac{R}{\ell_0}\right)^2 \right]^{1/(\gz_2-2\gk)}. 
\end{equation}
We may transform this relation into an equation by introducing a constant of proportionality via a critical Reynolds number $\cR^{(\gn)}_{2,uv}$ and write the leading-order estimate of $\ell_{uv}(R)$ as:
\begin{equation}
\frac{\ell_{uv}(R)}{\ell_0} = \left [ \frac{\cR^{(\gn)}_{uv}}{\cR^{(\gn)}_{2,uv}} \left( \frac{R}{\ell_0}\right)^2 \right]^{1/(\gz_2-2\gk)}. 
\end{equation}
Recall that this length scale marks the point $r=\ell_{uv}(R)$ where one enters the dissipation range in a correlation of two velocity differences where one of them is held constant at $R$ while the other is shrunk down to $r$. Implicit in our use of the fusion rules is the assumption that the two velocity differences have one point in common.  It is remarkable that the dissipation scale $\ell_{uv}(R)$ is an \emph{anomalous function} (i.e.. $R$ dependent) even for $F_2$.  This should be contrasted with the energy range of three-dimensional turbulence where the dissipation scale is an anomalous function only for the higher order correlations $F_n$ with $n>2$. \cite{article:Procaccia:1996:2}.

In the hyperdiffusion case $\gk >1$, we have $\gz_2-2\gk <0$, when the hypothetical intermittency corrections respect Eyink's constraint $\gz_2 < 11/3$ \cite{article:Eyink:1995}.  It follows that with increasing Reynolds number the separation of scales increases with the dissipation scale $\ell_{uv}(R)$ approaching 0. When $\gk <1$, with increasing Reynolds number, the dissipation  scale $\ell_{uv}(R)$ diverges to large scales instead of small scales, thereby demonstrating that such terms cannot provide a dissipation sink at small scales.  The case $\gk =1$ with $\gz_2 >2$ is interesting because it does not provide a well-behaved function either, and see below for more comments.  Finally,  when $\gk =1$ and $\gz_2 =2$, this evaluation is not valid, unless the logarithmic correction is taken into account. This case is discussed in the next subsection.

It should be remembered that the dissipation scale $\ell_{uv}(R)$ is not observed in the energy spectrum.  The standard dissipation scale that we do observe can be found by solving the equation $\ell_{uv}(\gl_{uv}) = \gl_{uv}$.  This leads to, 
\begin{equation}
\left(\frac{\gl_{uv}}{\ell_0}\right)^A = \left [ \frac{\cR^{(\gn)}_{uv}}{\cR^{(\gn)}_{2,uv}}  \right]^{1/(\gz_2-2\gk)}, 
\end{equation}
where the scaling exponent $A$ is given by
\begin{equation}
A =1-\frac{2}{\gz_2 -2\gk} = \frac{\gz_2 -2(\gk+1)}{\gz_2 -2\gk}
\end{equation}
and the solution is
\begin{equation}
\left(\frac{\gl_{uv}}{\ell_0}\right)= \left [ \frac{\cR^{(\gn)}_{uv}}{\cR^{(\gn)}_{2,uv}}  \right]^{1/(\gz_2-2(\gk+1))}. 
\end{equation}
When $\gk >1$, the separation of scales between this dissipation scale $\gl_{uv}$ and the integral length scale $\ell_0$ still increases with increasing Reynolds number, as we expect it to.  For the case $\gk =1$ and $\gz_2 =2$, this evaluation is identical to the dissipation scale evaluation that can be obtained by dimensional analysis.  However,  as we have noted in the previous paragraph, the function $\ell_{uv}(R)$, from which this evaluation has been obtained, is not valid in the absense of a logarithmic correction.

The dissipation scale function  $\ell_{uv}(R)$ is admissible if it satisfies the condition 
\begin{equation}
a\gl_{uv} > \ell_{uv}(a\gl_{uv}) \; ,\forall a>1.
\end{equation}
The condition holds if and only if 
\begin{align}
\frac{\ell_{uv}(a\gl_{uv})}{a\gl_{uv}} & =\frac{a^{2/(\gz_2-2\gk)}\ell_{uv}(\gl_{uv}) }{a\gl_{uv}}\\
&=a^{-1+2/(\gz_2-2\gk)} = a^{-A} < 1,\; \forall a>1,
\end{align}
which is true if and only if 
\begin{equation}
A =\frac{\gz_2 -2(\gk+1)}{\gz_2 -2\gk}>0.
\end{equation}
Note that the same condition guarantees that $\gl_{uv}$ goes to zero when the Reynolds number is taken to infinity. For $\gk>1$ and $\gz_2<11/3$, it is easy to see that $\gz_2-2(\gk+1) < \gz_2-4 < 11/3-4=-1/3<0$, therefore, the condition $A>0$ requires that $\gz_2-2\gk <0$.

The remarkable result is that in the physically relevant case $\gk =1$, the constraint imposed on any hypothetical intermittency corrections is $\gz_2>4$ (steeper than  $k^{-5}$).  Because it contradicts the constraint $\gz_2 <11/3$ of Eyink \cite{article:Eyink:1995}, we conclude that \emph{we may not have an enstrophy inertial range with intermittency corrections when $\gk =1$}.   In exact terms, we have shown that an enstrophy cascade with intermittency corrections will be destabilized if it is dissipated by standard molecular diffusion.  It is an interesting coincidence, and one that may warrant some further reflection, that the slope $k^{-5}$ also occurs in the Tran-Bowman theory \cite{article:Shepherd:2002,article:Bowman:2003,article:Bowman:2004} at the downscale range, where the two-dimensional Navier-Stokes equation does not have an infrared sink.

Although intermittency corrections do not seem to be forbidden in the hyperdiffusion case, as far as stability with respect to dissipation is concerned, a more careful study \cite{article:Eyink:2001} has already revealed additional constraints that exclude the intermittency scenario altogether, for the case of an enstrophy cascade with asymptotically constant enstrophy flux.  Numerical simulations of the enstrophy cascade also corroborate the absense of intermittency corrections \cite{article:Alvelius:2000,article:Falkovich:2002,article:Xiao:2003}.   Nevertheless, this analysis sheds some light into the fundamental differences between an enstrophy range dissipated by ordinary diffusion and one dissipated by hyperdiffusion.  In the hyperdiffusion case, an enstrophy range that has  already been somewhat destabilized into a steeper slope by the presense of a downscale energy flux \cite{article:Gkioulekas:2008:1} will not be further   disturbed by the dissipation range.  In the ordinary diffusion case, we may find it more difficult to obtain a robust enstrophy cascade with $k^{-3}$ scaling, because any steepening deviation from that scaling caused by perturbation from the forcing term $Q_n$ or the sweeping term $I_n$ would provoke further disturbance from the dissipation range.  This is why it is very significant to investigate numerically whether an enstrophy cascade can exist under regular diffusion.

We will now consider the dissipation scale functions for the higher order correlations $F_n$.  As usual, one velocity difference is shrinked down to scale $r$ while the others are kept constant at scale $R$.  It is further assumed that the small velocity difference is sharing an endpoint with at least one of the other velocity differences.  Let us assume that the boundary between the enstrophy range and the dissipation range is located on the line $r=\ell_{uv}^{(n)}(R)$.  To find the function $\ell_{uv}^{(n)}$, we start from the matching condition 
\begin{equation}
\frac{F_2 (\ell_{uv}^{(n)}(R))F_n (R)}{F_2 (R)}\sim \frac{(\ell_{uv}^{(n)}(R))^{2\gk}F_{n+1}(R)}{\nu R}.
\end{equation}
and solve for $\ell_{uv}^{(n)}(R)$. A simple calculation, given in detail in Appendix \ref{app:enstr-with-int}, gives the leading-order estimate of $\ell_{uv}^{(n)}(R)$ as::
\begin{equation}
\frac{\ell_{uv}^{(n)}(R)}{\ell_0} = \left( \frac{R}{\ell_0}\right)^{x_n} \left [ \frac{\cR^{(\gn)}_{uv}}{\cR^{(\gn)}_{n,uv}}  \right]^{1/(\gz_2-2\gk)},
\end{equation}
where the scaling exponents $x_n$ are given by 
\begin{equation}
x_n = \frac{\gz_{n+1}-\gz_n+\gz_2 -1}{\gz_2-2\gk}. 
\end{equation}
Note that the use of  matching conditions, corresponding to different values of $n$, may be introducing a different constant of proportionality, consequently the critical Reynolds numbers $\cR^{(\gn)}_{n,uv}$ will depend on $n$.  These higher order dissipation scale functions have the same dependence on the Reynolds number as $\ell_{uv}(R)$.  So, in that respect they behave similarly.  The only difference is that the anomalous correction now scales with $x_n$.

The standard dissipation scales for $F_n$ are found by solving the equation $\ell_{uv}^{(n)}(\gl_{uv}^{(n)})= \gl_{uv}^{(n)}$, leading to
\begin{equation}
\frac{\gl_{uv}^{(n)}}{\ell_0} = \left [\frac{\cR^{(\gn)}_{uv}}{\cR^{(\gn)}_{n,uv}}\right]^{1/[1-2\gk-(\gz_{n+1}-\gz_n)]}.
\end{equation}
Details are given in Appendix \ref{app:enstr-with-int}. Note that as the Reynolds number goes to infinity, the dissipation scale $\gl_{uv}^{(n)}$ goes to zero, as it is supposed to, even for the case $\gk=1$. Still, that does not necessarily mean that all these scenarios are admissible. The inertial range of the higher order correlations is admissible if it satisfies. 
\begin{equation}
\frac{\ell_{uv}^{(n)}(a\gl_{uv}^{(n)})}{a\gl_{uv}^{(n)}} =\frac{a^{x_n}\ell^{(n)}_{uv}(\gl_{uv}^{(n)})}{a\gl_{uv}^{(n)}} =a^{x_n-1} < 1,\; \forall a>1.
\end{equation}
This happens if and only if $(x_n -1) \ln a < 0 ,\; \forall a>1,$ and that requires $1-x_n >0$. Note that $\gz_{n+1}-\gz_n>0$, by the \Holder inequalities for a downscale cascade \cite{article:Gkioulekas:2008:1}, which implies that $(1-x_n)(\gz_2-2\gk) = 1-2\gk-(\gz_{n+1}-\gz_n) < 1-2\gk < 0, \forall \gk \geq 1$. It is therefore necessary that $\gz_2 -2\gk<0$. 
For the case, $\gk =1$ the constraint $\gz_2 -2\gk<0$ can never be satisfied when $\gz_2>2$, and that again excludes  intermittency corrections.

\subsection{Enstrophy range without intermittency corrections}

We now repeat the previous analysis for the case $\gz_n=n$ and $\gk=1$. This particular case deserves special attention, because in this case, and this case only, the leading dependence of $\ell_{uv}^{(n)}(R)$  on the Reynolds number $\cR^{(\gn)}_{uv}$  becomes exponential. Note that when $\gk>1$, the evaluation of the dissipation scales in the intermittency case above also applies to case $\gz_n = n$ with subleading corrections introduced by the logarithmic correction to the structure functions $F_n$ which, to first approximation, we can safely ignore.  When $\gk=1$, on the other hand, the role of these logarithmic corrections is crucial and has to be taken into account.  

We use the following evaluation for the correlations
\begin{equation}
F_n (R) \sim (\gn_{uv}^{1/3}R)^n [\ln (\ell_0/R)]^{a_n},
\end{equation}
 where $a_n$ are the scaling exponents of the logarithmic correction.  The prediction of Falkovich and Lebedev \cite{article:Lebedev:1994,article:Lebedev:1994:1} is that $a_n =2n/3$.  To first approximation, we have disregarded the Bowman correction \cite{article:Bowman:1996} to the logarithmic factor, which is negligible  when $R$ is not close to $\ell_0$, in terms of order of magnitude. The balance condition for the dissipation length scale of $F_2$ leads to
\begin{equation}
(\gn_{uv}^{1/3}\ell_{uv}(R))^2 [\ln (\ell_0/\ell_{uv}(R))]^{2/3}\sim \ell_{uv}(R)^2 \frac{\gn R^3}{\nu R} [\ln (\ell_0/R)]. 
\end{equation}
 Note that without the logarithmic correction, the dissipation scale function $\ell_{uv}(R)$ cancels out completely from the matching condition.  This means that the condition is always satisfied, and therefore a hypothetical $k^{-3}$ spectrum without logarithmic correction  would have to be part of the dissipation range for all wavenumbers $k$.  The presence of the logarithmic correction however gives a solution for the dissipation scale  $\ell_{uv}(R)$. Introducing a critical Reynolds number $\cR^{(\gn)}_{2,uv}$, we may therefore write the leading-order estimate of $\ell_{uv}(R)$ as:
\begin{equation}
\frac{\ell_{uv}(R)}{\ell_0} = \exp\left [ - \fracp{\cR^{(\gn)}_{uv}}{\cR^{(\gn)}_{2,uv}}^{3/2} \fracp{R}{\ell_0}^{3} \left [\ln\fracp{\ell_0}{R}\right]^{3/2} \right ].
\end{equation}
Note the exponential dependence of $\ell_{uv}(R)$ on the Reynolds number $\cR^{(\gn)}_{uv}$. 
In the limit of infinite Reynolds number, the dissipation scale $\ell_{uv}(R)$ goes to zero, thus it is well-behaved.  Solving the equation $\ell_{uv}(\gl_{uv})=\gl_{uv}$, the standard dissipation scale $\gl_{uv}$ is found to satisfy 
\begin{equation}
\fracp{\ell_0}{\gl_{uv}}^{3} \left [\ln\fracp{\ell_0}{\gl_{uv}}\right ]^{-1/2} = \fracp{\cR^{(\gn)}_{uv}}{\cR^{(\gn)}_{2,uv}}^{3/2}.
\end{equation}
 This equation cannot be solved in closed form, however  when the separation of scales is large we get
 \begin{equation}
\fracp{\ell_0}{\gl_{uv}}\approx\fracp{\cR^{(\gn)}_{uv}}{\cR^{(\gn)}_{2,uv}}^{1/2}, 
\end{equation}
consistent with the evaluation obtained from dimensional analysis.  

In general, the dissipation scale functions $\ell_{uv}^{(n)}(R)$ of the correlations $F_n$ can be derived directly from the matching condition
\begin{equation}
\frac{F_2 (\ell_{uv}^{(n)}(R))F_n (R)}{F_2 (R)}\sim \frac{(\ell_{uv}^{(n)}(R))^{2\gk}F_{n+1}(R)}{\nu R}.
\end{equation}
Solving for $\ell_{uv}^{(n)}(R)$ we find, for the case of generalized, as opposed to Falkovich-Lebedev scaling, that the leading-order estimate is:
\begin{equation}
\frac{\ell_{uv}^{(n)}(R)}{\ell_0}=
\exp\left [ - \fracp{\cR^{(\gn)}_{uv}}{\cR^{(\gn)}_{2,uv}}^{1/a_2}
\fracp{R}{\ell_0}^{2/a_2}
\left [\ln\fracp{\ell_0}{R}\right]^{b_n}\right ],
\end{equation}
where the logarithmic scaling exponents $b_n$ are  given by 
\begin{equation}
b_n = \frac{a_{n+1}-a_n+a_2}{a_2}.
\end{equation}
The details are given in appendix \ref{app:enstr-without-int}. We stress again that this result is confined only to the special case $\gz_n=n$ and $\gk=1$ in which the logarithmic factors are dominant, as discussed previously. For Falkovich-Lebedev scaling \cite{article:Lebedev:1994,article:Lebedev:1994:1}, the scaling exponent $a_n$ reads  $a_n = 2n/3$, which gives $b_n=3/2$ for all $n$. For $n=2$ and $a_n = 2n/3$, the equation for $\ell_{uv}^{(n)}(R)$ reduces to our previous equation for $\ell_{uv}(R)$. 

The standard dissipation scale $\gl_{uv}^{(n)}$ is the solution of the equation $\ell_{uv}^{(n)} (\gl_{uv}^{(n)}) = \gl_{uv}^{(n)}$ which simplifies to the following transcendental equation 
\begin{equation}
\fracp{\ell_0}{\gl_{uv}^{(n)}}^{2/a_2}
\left [\ln \fracp{\ell_0}{\gl_{uv}^{(n)}}\right]^{1-b_n} = \fracp{\cR^{(\gn)}_{uv}}{\cR^{(\gn)}_{2,uv}}^{1/a_2},
\end{equation}
and in the limit of large Reynolds numbers it vanishes as $\cR^{-1/2}$. 

 We can now confirm that the dissipation scale function $\ell_{uv}^{(n)}(R)$ satisfies the admissibility condition $\ell_{uv}^{(n)} (a\gl_{uv}^{(n)}) < a\gl_{uv}^{(n)}$ for all $a\gl_{uv}^{(n)} \in (\gl_{uv}^{(n)}, \ell_0)$. The argument is somewhat tedious, but given in detail in appendix \ref{app:enstr-without-int}. We find that the admissibility condition is satisfied if  $\gl_{uv}^{(n)} < \ell_0 \exp (-b_n a_2/2)$. This condition is readily satisfied with sufficient separation between $\gl_{uv}^{(n)}$ and $\ell_0$. For Falkovich-Lebedev scaling, we have $a_2=2/3$ and $b_n=3/2$ and the condition reduces to $\gl_{uv}^{(n)} < \ell_0 / \sqrt{e}$.  It should be noted that the proof in appendix \ref{app:enstr-without-int}  uses the claim that $\ell_{uv}^{(n)}(\ell_0)=\ell_0$ which requires in turn that  $b_n >0$. In general, utilizing only mathematical, as opposed to physical considerations, it can be shown that the assumption that the logarithmic scaling exponents $a_2$ and $a_3$ satisfy  $a_2>0$ and $a_3>0$ implies that $b_n > 0$ for all $n \in \bbN$ with $n>1$ (see appendix \ref{app:inequalities}).

We conclude that in the physical case $\gz_n=n$ and $\gk =1$, Kraichnan scaling is admissible but it is also necessary to have a logarithmic correction to allow the inertial range to form.  In effect, the role of the logarithmic correction is to ``inflate'' the region between the  curve $r=\ell_{uv}^{(n)}(R)$ and  the line $r=R$.  In the hypothetical case where the logarithmic correction is absent, the two curves  will coincide and that won't leave any room for local interactions. 

 It is interesting to note that when we have hyperdiffusion, the logarithmic correction is not needed to inflate the inertial range region. This points to   an interesting difference between the case of physical diffusion and hyperdiffusion. In the hyperdiffusion case, the inertial range will be reasonably inflated as soon as the energy spectrum has nearly converged to $k^{-3}$ scaling, and we may expect that it will exhibit inertial range behavior before it converges completely.  In the case of physical diffusion,  on the other hand, the enstrophy range will not begin to inflate until it  begins converging toward the logarithmic correction.  In fact, it is possible that the absence of a region in the $(r,R)$ plane in which local interactions can occur, when the spectrum is still steep, might make it impossible for the enstrophy range to converge at all.

\section{The case of the inverse energy cascade}

For the inverse energy cascade, the dissipation scale theory is a lot simpler. We let $\gee_{ir}$ represent  the upscale energy flux. We begin with the evaluation given by the $3/2$ law \cite{article:Tabeling:1998,article:Bernard:1999,article:Bernard:2000,article:Davidson:2008}, that $F_3 (r) \sim \gee_{ir} r$, and for the other correlations we use 
\begin{equation}
F_n (r) \sim (\gee_{ir} r)^{n/3} (r/\ell_0)^{\gz_n -n/3}. 
\end{equation}
The dissipation scale is obtained from the matching condition
\begin{equation}
\frac{F_2 (\ell_{ir}^{(n)}(r))F_n (r)}{F_2 (r)}\sim \frac{(\ell_{ir}^{(n)}(r))^{-2m}F_{n+1}(r)}{\gb r},
\label{eq:invenmatchcond}
\end{equation}
 and that leads for $n=2$ to 
\begin{equation}
(\gee_{ir}\ell_{ir}(r))^{2/3} (\ell_{ir}(r)/\ell_0)^{\gz_2-2/3}
\sim \frac{\ell_{ir}(r)^{-2m}\gee_{ir} r}{\gb r}.
\end{equation} 
It follows that $\ell_{ir}(r)$ is constant with respect to $r$ with $\ell_{ir}(r)=\ell_{ir}$, and the dissipation scale $\ell_{ir}$  reads, to leading order:
\begin{equation}
\frac{\ell_{ir}}{\ell_0} =\fracb{\cR^{(\gee)}_{ir}}{\cR^{(\gee)}_{2,ir}}^{1/(\gz_2+2m)},
\end{equation}
where $\cR^{(\gee)}_{ir}$ is the Reynolds number  corresponding to the inverse energy cascade given by
\begin{equation}
\cR^{(\gee)}_{ir} =\frac{\gee_{ir}^{1/3}\ell_0^{-2/3+2m}}{\gb}.
\end{equation}
This dissipation scale corresponds to the partially fused correlation $F_2$ where one of the two velocity differences is being stretched towards larger scales while the other velocity difference is held fixed at  $r$.  Note that, unlike the case of the enstrophy cascade, here the dissipation scale $\ell_{ir}(r)$ is independent of $r$. Consequently, for all $r$ in the inertial range, $\ell_{ir}(r)$ coincides with the standard dissipation scale $\gl_{ir}$, corresponding to the cross-over to the dissipation range  when \emph{both} velocity differences are stretched simultaneously, i.e. $\ell_{ir}(r) = \gl_{ir}$.  We have a similar situation in the energy cascade of three-dimensional turbulence \cite{article:Procaccia:1996:2}.  In the limit of the Reynolds number going to infinity, the dissipation scale  $\ell_{ir}$ also goes to infinity because $\gz_2+2m>0$.  It is interesting to note that this also holds when $m=0$.  In other words, Ekman damping can provide a sink for the inverse energy cascade, whereas we've seen  earlier that it cannot provide one for the enstrophy cascade. 

The dissipation scales that correspond to the correlations $F_n$ with $n>2$, on the other hand, are functions of $r$.  Again, we consider the case where the velocity differences are partially fused, and one velocity difference is being stretched to large scales while all others remain constant at $r$, and we write the corresponding dissipation scale as $\ell_{ir}^{(n)}(r) = \ell_{ir} (r/\ell_0)^{x_n}$.  From this, we may rewrite $F_2 (\ell_{ir}^{(n)}(r))$ as:
\begin{equation}
\begin{split}
F_2 (\ell_{ir}^{(n)}(r))
&\sim
F_2 (\ell_{ir})(r/\ell_0)^{x_n\gz_2}
\sim
\frac{F_3 (r)}{\gb r \ell_{ir}^{2m}}
(r/\ell_0)^{x_n\gz_2}\\
&=
\frac{F_3 (r)}{\gb r (\ell_{ir}^{(n)}(r))^{2m}}
(r/\ell_0)^{x_n\gz_2 + 2x_n m}.
\end{split}
\end{equation}
Substituting that into the matching condition \eqref{eq:invenmatchcond} reads
\begin{equation}
\frac{F_{n+1}(r)}{\gb r (\ell_{ir}^{(n)}(r))^{2m}}
\sim 
\frac{F_3 (r)}{\gb r (\ell_{ir}^{(n)}(r))^{2m}}
(r/\ell_0)^{x_n\gz_2 + 2x_n m}\frac{F_n (r)}{F_2 (r)}.
\end{equation}
  After all the cancellations, we obtain the following constraint for the scaling exponents 
\begin{equation}
\gz_{n+1}-(n+1)/3 = (\gz_n -n/3) - (\gz_2-2/3)+x_n (\gz_2 +2m). 
\end{equation} 
Solving for $x_n$ yields 
\begin{equation}
x_n = \frac{\gz_{n+1}-\gz_n +\gz_2 -1}{\gz_2 + 2m}.
\end{equation}
Note that in an inverse cascade $\gz_{n+1}-\gz_n$ forms an increasing sequence, by the \Holder inequalities \cite{article:Gkioulekas:2008:1}, and since $x_2 = 0$, it follows that $x_n$ is an increasing sequence and therefore
\begin{equation}
x_n \ge 0,\;\forall n >2.  
\end{equation}
Next, we solve the equation $\ell_{ir}^{(n)}(\gl_{ir}^{(n)})=\gl_{ir}^{(n)}$, and obtain the standard dissipation scales, which read, to leading order:
\begin{equation}
\gl_{ir}^{(n)} =\fracp{\gl_{ir}}{\ell_0^{x_n}}^{1/(1-x_n)}.
\end{equation}  The dissipation scale function itself is admissible if it satisfies the condition
\begin{equation}
\ell_{ir}^{(n)}(a \gl_{ir}^{(n)})=
a^{x_n}\ell_{ir}^{(n)}(\gl_{ir}^{(n)})= 
a^{x_n}\gl_{ir}^{(n)} > 
a\gl_{ir}^{(n)},\;\forall a<1,
\end{equation}  which holds if and only if
\begin{equation}
(1-x_n)\ln a < 0,\;\forall a<1,
\end{equation} and this requires in turn that
\begin{equation}
1-x_n = \frac{(2m+1)-(\gz_{n+1}-\gz_n)}{\gz_2+2m} > 0,\;\forall n>2. 
\end{equation}
This condition will be satisfied as long as the scaling exponents satisfy the following constraint 
\begin{equation}
 \gz_{n+1}-\gz_n < 2m+1    ,\;\forall n>2. \label{eq:invintineq}
\end{equation}

From the \Holder inequalities, we know that $\gz_{n+1}-\gz_n$ is either constant (for the case of no intermittency corrections) or increasing (otherwise) \cite{article:Gkioulekas:2008:1}. Therefore,  in the presense of hypothetical intermittency corrections to the scaling exponents $\gz_n$, we should entertain the possibility that this inequality could be violated for large enough $n$. If that were to happen, then there would be no possibility of an inertial range for generalized structure functions $F_n$ for large values of $n$, which would make for an interesting situation indeed. Here's what we can say with certainty: To violate this inequality for the least favorable case $m=0$, we require multifractal contributions to the inverse energy cascade with \Holder exponents $h>1$. In our previous paper  \cite{article:Gkioulekas:2008:1}, we have shown that such contributions do \emph{not} necessarily violate locality in the $\cO_n F_{n+1}$ term. Furthermore, they do \emph{not} violate the stability constraint $h \geq 1/3$ with respect to gaussian forcing either! If $\gz_n$ grows linearly as $n\to +\infty$, then there exists an $m$ such that the inequality can hold for all $n$. If, however, the growth rate of $\gz_n$ is faster than linear, then for any choice of $m$ there is  a $n_0$ for which the inequality is  violated for all $n>n_0$.  Consequently,  as long as the question of whether the inverse energy cascade has intermittency corrections to the scaling exponents $\gz_n$ with superlinear growth rate remains open, the question of whether the inequality \eqref{eq:invintineq} is broken at large $n$ also remains open.

Obviously, a possible violation of the inequality \eqref{eq:invintineq}, were it to occur,  can be ameliorated by increasing $m$. On the other hand,  it is reasonable to expect that a receding dissipation profile, that comes with increasing $m$, would eventually fail to dissipate the sweeping term effectively \cite{article:Gkioulekas:2007}, leading to the non-universal behaviour of Danilov-Gurarie \cite{article:Gurarie:2001,article:Gurarie:2001:1,article:Danilov:2003}. Thus, the realization of a steady-state inverse energy cascade could require a  balancing act in setting up the dissipation sink: a  low-order $m$ dissipates the sweeping term $I_n$ more efficiently but makes it easier to break the inequality \eqref{eq:invintineq}. A higher-order $m$ will delay the violation of \eqref{eq:invintineq} considerably, but at the price of less efficient dissipation of the sweeping effect.

\section{On the existence of anomalous sinks} 

In this section we will now use some of the above results to show that if the fusion rules are satisfied then the corresponding cascades will have anomalous sinks. Furthermore, we will show that if the fusion rules are not satisfied, then  the dissipation scales will not be positioned in a manner that  provides for anomalous sinks. Therefore, in cases where either the enstrophy cascade or the inverse energy cascade have been observed experimentally or numerically, the fusion rules hypothesis has to hold.



Before we proceed with the main argument, we have to first take care of the following difficulty: In the argument of the preceding sections we have made the  casual assumption that we may evaluate the generalized structure function $F_3$ using  the $1/8$-law for the  enstrophy cascade or the  $3/2$-law   for the inverse energy cascade.  The immediate concern is that we could be engaging in a circular argument if the anomalous sink hypothesis is needed to establish these laws in the first place. Fortunately, this does not affect an argument by contradiction where we deliberately assume the existence of anomalous sinks and the violation of the fusion rules to deduce a contradiction. On the other hand, we would like to have a way to break this apparent  vicious circle.

In subsection A we will explain how this can be done. In subsection B we give the main argument itself. The reader who wishes to skip ahead to the main argument can continue reading from subsection B.


\subsection{The logical structure of  the argument}

Let us begin by considering, with no loss of generality, the case of the enstrophy cascade. The validity of the $1/8$-law within an appropriate interval of length scales requires the following three conditions:
(1) most of the enstrophy must flow downscale and most of the energy must flow upscale;  
(2) there must be a corresponding inertial range situated within a wide separation of scales between the forcing scale and the dissipation scale;  
(3)  the positioning of the dissipation scale must be governed by a law such that for fixed viscosity it can dissipate any arbitrary amount of energy and enstrophy input.  
A similar set of conditions are needed for the validity of the 3/2 law, for the case of the inverse energy cascade. We will now argue that the first condition can be established from first principles. As for the second condition, I will suggest that it is possible to  deal with it provided that we adopt a more careful interpretation of the argument of the previous two sections. 


The first condition, namely that  most of the enstrophy must be constrained to flow downscale while  most of the energy  must be constrained to  flow upscale, can be  established from first principles, without assuming a priori  the existence of anomalous sinks. The original argument, proffered  as proof establishing this claim, was given a long time ago by  \Fjortoft \cite{article:Fjortoft:1953}. It was later noticed \cite{article:Warn:1975,article:Welch:2001,article:Orlando:2003:1,article:Tung:2006} that the claim underlying his argument, that the twin conservation of energy and enstrophy  is the sole determining reason that constrains  the direction of fluxes in two-dimensional turbulence, is flawed. Instead, in a recent paper \cite{article:Tung:2007:1} co-authored with KK Tung, we have shown that it is the combined effect of the twin conservation laws and the mathematical structure of the dissipation operator that constrains energy to go mostly upscale and enstrophy to go mostly downscale.

Our key result, in that paper, is as follows: Let $\Pi_E (k)$ be the energy flux  and let $\Pi_G (k)$  be the enstrophy  flux transfered from the interval $(0,k)$ to $(k,+\infty)$ per unit time  by the nonlinear term in the Navier-Stokes equations.  If  the energy forcing spectrum $F_E (k)$ is confined to a narrow interval of wavenumbers $[k_1, k_2]$ such that $F_E (k) = 0$ when $k\not\in [k_1, k_2]$, then it can be shown for the forced-dissipative case, without making any ad hoc assumptions, that under stationarity, the energy flux $\Pi_E (k)$ and the enstrophy flux $\Pi_G (k)$ will satisfy the inequalities
\begin{align}
\int_0^k  q \Pi_E (q) \; \df{q} &<0,  \;\forall k > k_2, \label{eq:ineqE}\\
\int_k^{+\infty} q^{-3} \Pi_G (q)\; \df{q} &> 0, \;\forall k < k_1 \label{eq:ineqG}.
\end{align}
 The inequality \eqref{eq:ineqE} implies that the negative flux in the $(0,k_1)$ interval is more intense than the positive flux in the $(k_2, +\infty)$ because the weighted average of $\Pi_E (k)$ gives more weight to the large wavenumbers. Thus, \eqref{eq:ineqE}  implies that energy fluxes upscale in the net. Similarly, \eqref{eq:ineqG} implies that enstrophy fluxes downscale in the net. It follows that \emph{we may justify to some degree the evaluations used for $F_3$, in the sense that they represent the only self-consistent scenario in which inertial ranges exist. }

Unfortunately, this still leaves open the question of the successful existence of  inertial ranges where universal cascades are to be situated.   In  the argument of the previous two sections, we seem to assume without proof the existence of a forcing-dissipation configuration that allows cascades to exist. However, we can view the same argument in an entirely different context if we situate it within the broader argument, outlined in greater detail in the conclusion of the present paper, of which it is a part. 

The structure of the broader argument, in summary form, consists of 2 basic steps \cite{article:Tung:2005,article:Tung:2005:1,article:Gkioulekas:2008:1}: First, we show that the balance equations have two homogeneous solutions, corresponding to an energy cascade and an enstrophy cascade for both directions, and a particular solution driven by forcing and sweeping. The flux inequalities \eqref{eq:ineqE}  and \eqref{eq:ineqG} establish the downscale enstrophy cascade and the upscale energy cascade as the leading homogeneous solutions. Second, to show the conditions necessary for the existence of cascades with universal scaling,  we require the existence of a region $\cA_n \subseteq \bbR^{2n}$ where the homogeneous solutions dominate the particular solution  in the generalized structure function $F_n (\{\bfX\}_n,t)$ for all $\{\bfX\}_n \in \cA_n$ and for each choice of $n\in\bbN$ with $n>1$. A separate effect is the distortion of the homogeneous solutions by the dissipation term. This defines a separate region $\cB_n \subseteq \bbR^{2n}$ where this distortion effect is negligible. For the successful formation of a cascade we need to have a measurable  region $\cJ_n = \cA_n \cap \cB_n$ with non-zero measure wherein the regions $\cA_n$  and $\cB_n$ overlap. This overlap region $\cJ_n$ is a multidimensional representation of the extent of the inertial range for the corresponding cascade, whereby the leading homogeneous solution dominates the particular solution. 

The point is that both regions $\cA_n$  and $\cB_n$ are to be calculated separately first, and the existence of the cascade itself is \emph{then} decided by the existence of an overlap $\cJ_n$ between the two regions  $\cA_n$  and $\cB_n$. So, we can take the standpoint that the argument of the previous two sections is the determination of the region $\cB_n$,  whereas the argument of my previous papers \cite{article:Gkioulekas:2007,article:Gkioulekas:2008:1} concerns the determination of the region $\cA_n$. From this  standpoint, we can argue that \emph{in the previous two sections we were investigating the dissipative corrections to the homogeneous solutions independently of whether or not the homogeneous solutions actually dominate over the particular solution}. As we have shown previously \cite{article:Tung:2005,article:Tung:2005:1,article:Gkioulekas:2008:1}, the scaling exponent $\gz_3$ evaluations as $\gz_3 =1$ for an energy cascade and $\gz_3 =3$ for an enstrophy cascade are an inherent characteristic of the homogeneous solutions themselves that is unrelated entirely from the anomalous sink question.

\subsection{Anomalous sinks from the fusion rules}

With the above prefatory remarks under consideration, let us now proceed with the anomalous sink argument.  Let $E(k)$ be the energy spectrum, and suppose that a certain amount of energy and enstrophy is dissipated at large scales and another amount at small scales. The energy dissipation rates are given by the following integrals, 
\begin{equation}
\begin{split}
\gee_{ir} &= 2\gb\int_{0}^{+\infty} k^{-2m} E(k) \; \df{k},\\
\gee_{uv} &= 2\nu\int_{0}^{+\infty} k^{+2\gk} E(k) \; \df{k},
\end{split}
\end{equation}
and under steady state they satisfy $\gee = \gee_{ir} + \gee_{uv}$, where $\gee$ is the rate of energy input. Here, $\gee_{ir}$ is the energy dissipation rate at large scales and $\gee_{uv}$ is the energy dissipation rate at small  scales.  Similarly, the enstrophy dissipation rates are given by, 
\begin{equation}
\begin{split}
\gn_{ir} &= 2\gb\int_{0}^{+\infty} k^{-2m+2} E(k) \; \df{k},\\
\gn_{uv} &= 2\nu\int_{0}^{+\infty} k^{+2\gk+2} E(k) \; \df{k},
\end{split}
\end{equation}
and, again, under steady state they satisfy $\gn= \gn_{ir} + \gn_{uv}$, where $\gn$ is the rate of enstrophy input. Here, $\gn_{ir}$ is the enstrophy dissipation rate at large scales and $\gn_{uv}$ is the enstrophy dissipation rate at small  scales.
 
We begin with the case of the inverse energy cascade. Since the effect of the sink at small scales can be safely ignored at large scales, the dominant contribution to the  energy  dissipation rate is given by the integrals 
\begin{align}
\gee_{ir} &= 2\gb\int_{0}^{1/\gl_{ir}} k^{-2m} E(k) \; \df{k} + 2\gb\int_{1/\gl_{ir}}^{1/\ell_0} k^{-2m} E(k) \; \df{k}\\ & \quad + 2\gb\int_{1/\ell_0}^{+\infty} k^{-2m} E(k) \; \df{k}.
\end{align}
Here $\gl_{ir}$ is the standard dissipation scale at large scales. For the case of the inverse energy cascade we have found that $\ell_{ir} (r)$ is independent of $r$ with $\ell_{ir} (r) = \gl_{ir}$. We have also shown that
\begin{equation}
\frac{\gl_{ir}}{\ell_0} =\fracb{\cR^{(\gee)}_{ir}}{\cR^{(\gee)}_{2,ir}}^{1/(\gz_2+2m)} \sim  \gb^{-1/(\gz_2+2m)}.
\end{equation}
The key point is that the scaling exponent $\gz_2$ in the above equation originates from the order $R$ velocity difference separation in the $(n,p)=(2,1)$  fusion rule, which generally scales as $R^{\gz_2-\xi_{2,1}}$. The fusion rule hypothesis gives $\xi_{2,1}=0$ leading to the above estimate of $\gl_{ir}$. If this fusion rule were to be violated, the scaling exponent $\gz_2-\xi_{2,1}$ would replace $\gz_2$, and the dissipation scale $\gl_{ir}$ would then follow
\begin{equation}
\gl_{ir}\sim  \gb^{-1/(\gz_2-\xi_{2,1}+2m)}.
\end{equation}
In the limit of extending the separation of scales in the inverse energy cascade, we have $\ell_0\to 0^{+}$ which kills the third integral and $\gb\to 0$ which gives $\gl_{ir}\to +\infty$ which kills the first integral. It follows that the dominant contribution comes from the second integral. When we substitute $E(k)\sim  k^{-(1+\gz_2)}$ to the second integral we get:
\begin{align}
\gee_{ir} &\sim \gb \int_{1/\gl_{ir}} k^{-2m}  k^{-(1+\gz_2)} \; \df{k} \sim \gb (1/\gl_{ir})^{-(\gz_2+2 m)} \\ &\sim \gb \gl_{ir}^{\gz_2+2 m} \sim  \gb^{1-(\gz_2+2 m)/(\gz_2-\xi_{2,1}+2m)}.
\end{align}
It follows that \emph{the anomalous sink hypothesis for the inverse energy cascade which requires that $\gee_{ir}$ be independent of $\gb$ is satisfied if and only if the fusion rules hypothesis $\xi_{2,1}=0$ holds.} In other words, we cannot have an anomalous sink if the fusion rules hypothesis is violated. It is interesting to note that  large $m$ has a tendency to mitigate the impact on the sink anomaly of a small discrepency between $\gz_2-\xi_{2,1}$ and $\gz_2$. That is, when $m\to +\infty$, $\gee_{ir}$ becomes independent of $\gb$ regardless of the value of the scaling exponent $\xi_{2,1}$.

A similar argument can be employed to establish the existence of anomalous enstrophy sink at small scales. The dominant contribution to the enstrophy flux is given by the integrals 
\begin{align}
\gn_{uv} &= 2\nu \int_{0}^{1/\ell_0} k^{2\gk+2} E(k) \; \df{k} +  2\nu \int_{1/\ell_0}^{1/\gl_{uv}} k^{2\gk+2} E(k) \; \df{k} \\
\quad &+ 2\nu \int_{1/\gl_{uv}}^{+\infty} k^{2\gk+2} E(k) \; \df{k}.
\end{align}
Here $\gl_{uv}$ is the standard dissipation scale at small scales. For the case of the enstrophy cascade with hyperdiffusion (i.e. $\gk>1$) we have found that $\ell_{uv}(R)$ is anomalous, in the sense that it is $R$ dependent and reads
\begin{equation}
\frac{\ell_{uv}(R)}{\ell_0} = \left [ \frac{\cR^{(\gn)}_{uv}}{\cR^{(\gn)}_{2,uv}} \left( \frac{R}{\ell_0}\right)^2 \right]^{1/(\gz_2-2\gk)}. 
\end{equation}
As long as $\gz_2-2\gk <0$, the standard dissipation  scale $\gl_{uv}$ is the solution of  the equation $\ell_{uv}(\gl_{uv}) = \gl_{uv}$ and it reads:
\begin{equation}
\left(\frac{\gl_{uv}}{\ell_0}\right)= \left [ \frac{\cR^{(\gn)}_{uv}}{\cR^{(\gn)}_{2,uv}}  \right]^{1/(\gz_2-2(\gk+1))}  \sim  \nu^{-1/(\gz_2-2(\gk+1))}. 
\end{equation}
Again, the scaling exponent $\gz_2$ now originates from the order $r$ velocity difference separation in the $(n,p)=(2,1)$  fusion rule, which  scales as $r^{\xi_{2,1}}$, consequently, under violation of the fusion rules hypothesis, the scaling exponent $\xi_{2,1}$ would replace $\gz_2$, and the dependence of $\gl_{uv}$ on $\nu$ would be given by 
\begin{equation}
\gl_{uv} \sim  \nu^{-1/(\xi_{2,1}-2(\gk+1))}.
\end{equation}
Now, by repeating a similar line of  argument as in the previous case of the inverse energy cascade, we argue that in  the limit of extending the separation of scales in the  enstrophy cascade the dominant contribution to $\gn_{uv}$  comes from the second integral. Substituting the enstrophy spectrum $E(k)\sim  k^{-(1+\gz_2)}$ to the second integral gives:
\begin{align}
\gn_{uv} &\sim  \nu \int^{1/\gl_{uv}} k^{2\gk+2} k^{-(1+\gz_2)} \; \df{k} \sim \nu (1/\gl_{uv})^{2\gk+2-\gz_2} \\
&\sim \nu \gl_{uv}^{\gz_2-2(\gk+1)}
\sim  \nu^{1-(\gz_2-2(\gk+1))/(\xi_{2,1}-2(\gk+1))}.
\end{align}
Consequently, we find again that \emph{the anomalous sink hypothesis for the enstrophy cascade which requires that $\gn_{uv}$ be independent of $\nu$ is satisfied if and only if the fusion rules hypothesis $\xi_{2,1}=\gz_2$ holds.} Again we see that large $\gk$ hyperdiffusion helps the enstrophy sink to remain approximately anomalous even under a small deviation of $\xi_{2,1}$ from $\gz_2$.

The careful reader will note that the evaluation of $\gl_{uv}$ used in the argument above is not valid when $\gk=1$ and $\gz_2=2$. We have already shown that in this case, $\gl_{uv}$ is given by the transcendental equation 
\begin{equation}
\fracp{\ell_0}{\gl_{uv}}^{3} \left [\ln\fracp{\ell_0}{\gl_{uv}} \right]^{-1/2} = \fracp{\cR^{(\gn)}_{uv}}{\cR^{(\gn)}_{2,uv}}^{3/2},
\label{eq:sepscales2}
\end{equation}
for the case of Falkovich-Lebedev scaling \cite{article:Lebedev:1994,article:Lebedev:1994:1}. It follows that 
\begin{equation}
(1/\gl_{uv})^{3} \sim \nu^{-3/2} [\ln (\ell_0/\gl_{uv})]^{1/2},
\end{equation}
and thus, repeating the calculation of $\gn_{uv}$, with the inclusion of the logarithmic correction to the energy spectrum $E(k) \sim  k^{-3} [\ln  (k\ell_0)]^{-1/3}$  now gives:
\begin{align}
\gn_{uv} &\sim \nu \int^{1/\gl_{uv}} k^{4} k^{-3} [\ln  (k\ell_0)]^{-1/3} \; \df{k}\\ 
&\sim   \nu (1/\gl_{uv})^{2} [\ln (\ell_0/\gl_{uv})]^{-1/3}\\
& \sim \nu [(1/\gl_{uv})^{3}]^{2/3}[\ln (\ell_0/\gl_{uv})]^{-1/3}  \\ 
&\sim \nu [\nu^{-3/2} [\ln (\ell_0/\gl_{uv})]^{1/2}]^{2/3}[\ln (\ell_0/\gl_{uv})]^{-1/3} \\
& \sim \text{ const },
\end{align}
which shows again that we have an anomalous sink. 

Incidentally, it is an interesting exercise to consider the case of arbitrary logarithmic scaling and derive the necessary and sufficient condition on the exponents $a_n$ for the existence of an anomalous energy sink. In general, the energy spectrum scales as $E(k) \sim k^{-3} [\ln  (k\ell_0)]^{a_2-1}$, and it follows that 
\begin{align}
\gn_{uv} &\sim \nu \int^{1/\gl_{uv}} k^{4} k^{-3} [\ln  (k\ell_0)]^{a_2-1} \; \df{k} \\
&\sim \nu (1/\gl_{uv})^{2} [\ln (\ell_0/\gl_{uv})]^{a_2-1}.
\end{align}
The transcendental equation for $\gl_{uv}$ now reads
\begin{equation}
\fracp{\ell_0}{\gl_{uv}}^{2/a_2} \left [\ln\fracp{\ell_0}{\gl_{uv}} \right]^{1-b_2} = \fracp{\cR^{(\gn)}_{uv}}{\cR^{(\gn)}_{2,uv}}^{1/a_2},
\end{equation}
which gives
\begin{equation}
\fracp{\ell_0}{\gl_{uv}}  \sim \nu^{-1/2}\left [\ln\fracp{\ell_0}{\gl_{uv}} \right]^{(b_2-1) a_2/2}.
\end{equation}
It follows that
\begin{align}
\gn_{uv} &\sim \nu \{\nu^{-1/2}  [\ln (\ell_0/\gl_{uv})]^{(b_2-1) a_2/2} \}^2 [\ln (\ell_0/\gl_{uv})]^{a_2-1}\\
 &\sim [\ln (\ell_0/\gl_{uv})]^{(b_2-1) a_2 + (a_2-1)} \sim [\ln (\ell_0/\gl_{uv})]^{a_2 b_2-1}.
\end{align}
Since $a_2 b_2-1 =  a_3-1$, we see that an anomalous enstrophy sink exists if and only if $a_3 = 1$. This is obviously consistent with the Falkovich-Lebedev claim \cite{article:Lebedev:1994,article:Lebedev:1994:1} that $a_n = 2n/3$.

 It is possible to combine this argument and the previous argument to generalize this result for the case $\gk > 1$, provided that the logarithmic corrections are taken into account in calculating the location of the dissipation scale $\gl_{uv}$. This gives the enstrophy dissipation rate $\gn_{uv}$  as:
 \begin{equation}
 \gn_{uv} \sim \nu^{1-(\gz_2-2(\gk+1))/(\xi_{2,1}-2(\gk+1))} [\ln (\ell_0/\gl_{uv})]^{a_3-1}.
 \end{equation}
Again, the fusion rules hypothesis rules out the power-law dependence of $\gn_{uv}$ on $\nu$. However, the condition $a_3 = 1$ is still needed to rule out a logarithmic dependence of  $\gn_{uv}$ on $\nu$. Future work should thus take another good look at these logarithmic scaling exponents. In connection with this remark, it is worth noting that, in recent papers \cite{article:Tran:2005,article:Dritschel:2006}, Tran claims that he proved that it is not possible for the enstrophy cascade to have an anomalous enstrophy sink. However, the range of applicability of his result is confined only to the case of freely decaying two-dimensional turbulence, with $\gk=1$ and no sink at  large scales (i.e. $\gb=0$). As such, it is not applicable under two sinks or for the forced-dissipative case.

Although the case of the energy cascade of three-dimensional turbulence is not the primary focus of this paper, it is easy to see that our argument can be repeated to derive the dependence 
\begin{equation}
\gee \sim \nu^{1-(\gz_2-2(\gk+1))/(\xi_{2,1}-2(\gk+1))},
\end{equation}
between the energy dissipation rate $\gee$  and the viscosity $\nu$.  Under the regular violation of the $(n,p)=(2,1)$ fusion rule, where $\xi_{2,1}$ satisfies $0<\xi_{2,1}<\gz_2$, we see that the energy dissipation rate $\gee$ will vanish  in the limit $\nu\to 0^{+}$, suggesting an insufficiently powerful energy sink. If and when such a violation does occur, the energy spectrum $E(k)$ might respond  by reducing the value of the scaling exponent $\gz_2$ until it converges to the value of $\xi_{2,1}$.  That would lead to a shallower slope for the energy spectrum $E(k)$. Such a response would not violate locality \cite{article:Gkioulekas:2008:1} and it is therefore reasonable to expect that the energy would still trickle down via local interactions and dissipate when dissipation is strong enough, thus maintaining  sink anomaly in the intermediate asymptotic sense. Our argument shows that \emph{a regular fusion rules violation combined with intermediate asymptotic sink anomally must lead to a shallower energy spectrum.} This is consistent with a bottleneck-type pile up of energy near the interface between the inertial range and the dissipation range. Thus, a question that should be considered  by future work is whether the well-known bottleneck problem \cite{article:Falkovich:1994:1,article:Matsumoto:2004} of  three-dimensional turbulence can be understood in terms of a regular violation of the fusion rules.

\section{Conclusion and discussion}

In this paper we have shown that there is a close relation between the fusion rules hypothesis and the anomalous sink hypothesis. This relation holds for both two-dimensional and three-dimensional turbulence, even though this paper was mainly focused on two-dimensional turbulence. We have also investigated the boundary between the inertial range and dissipation range for both inverse energy cascade and enstrophy cascade in terms of a two-dimensional representation instead of the usual one-dimensional interval-of-scales approach. The purpose of that investigation was to derive the cascade stability conditions with respect to the dissipation terms. These results add to the work presented in our previous paper \cite{article:Gkioulekas:2008:1} where we showed that, given the fusion rules hypothesis, we can investigate cascade locality  and  cascade stability with respect to the forcing terms. These results, when put together, are beginning to outline a fundamental, albeit yet incomplete, theory of two-dimensional turbulence. Because the details of the overall argument are quite technical, we will attempt in these concluding remarks to outline, in very broad strokes, the logical structure of the proposed theory. We will then review the plausibility of the fusion rules hypothesis, and conclude by listing some interesting insights that we have gained from the present paper.

 The point of departure are the generalized balance equations. \emph{First}, we observe that these equations have  a general solution that consists of two homogeneous solutions that represent correspondingly an enstrophy cascade with $\gz_3 = 3$ and an energy cascade with $\gz_3 = 1$ \cite{article:Tung:2005}, and a \emph{particular solution} driven by forcing and sweeping. The determination of the $\gz_3$ exponents for the two homogeneous solutions does not use the anomalous sink hypothesis. It follows instead from a self-consistency argument \cite{article:Tung:2005}. \emph{Second}, we derive the constraints \eqref{eq:ineqE}  and \eqref{eq:ineqG} on the fluxes $\Pi_E (k)$ and $\Pi_G (k)$, which imply that the only self-consistent manner in which these cascades can manifest themselves is as a downscale enstrophy cascade and an upscale inverse energy cascade \cite{article:Tung:2007:1}. It is possible, for the case of finite viscosities that side by side with the dominant cascades there may be a subleading downscale energy cascade and a subleading upscale enstrophy cascade \cite{article:Tung:2005}. However, a constraint on the permitted magnitude of the energy and enstrophy fluxes associated with these subleading cascades, prevents them from having a significant effect on the energy spectrum \cite{article:Tung:2005:1,article:Tung:2007}.

As was discussed in my previous papers \cite{article:Tung:2005,article:Tung:2005:1,article:Gkioulekas:2007,article:Gkioulekas:2008:1}, in order for either the inverse energy cascade or the enstrophy cascade to exist, there must be a region $\cA_n \subseteq \bbR^{2n}$ where the corresponding leading homogeneous solution dominates the particular solution driven by sweeping and forcing. This is the so-called \emph{stability condition}. Furthermore, the dissipation terms effectively act to distort the homogeneous solutions within a certain \emph{dissipative region}. If $\cB_n \subseteq \bbR^{2n}$ is the region where such dissipative effects on the leading homogeneous  solution are negligible, then our main requirement for the existence of a cascade is that there should be a measurable  overlap $\cJ_n = \cA_n \cap \cB_n$  between the two regions $\cA_n$  and $\cB_n$ with non-zero measure. The region $\cJ_n$ is thus a multidimensional representation of the extent of the inertial range associated with the generalized structure function $F_n$. 

We may now pose the following question: Does the structure of the mathematical theory of turbulence allow the existence of cascades with universal scaling? In other words, \emph{if we require the cascades to have universal scaling exponents, can the region $\cJ_n$ have a non-zero measure?}  The argument in response to the question now runs as follows: First, the demand that the cascades must have universal scaling implies that the cascades, if they exist, must satisfy the fusion rules hypothesis. As things stand now, this first step is the incomplete link in the overall chain, as far as two-dimensional turbulence is concerned. This is discussed further below.

Second, from the fusion rules hypothesis, we establish the locality of the local interactions term $\cO_n F_{n+1}$ of the balance equations and combined with the assumption of random gaussian forcing we then derive the stability conditions of either cascade with respect to forcing perturbations \cite{article:Gkioulekas:2008:1}. The question of stability with respect to sweeping is still an open problem \cite{article:Gkioulekas:2007}. Both kinds of stability are necessary conditions for the existence of cascades with universal scaling, because if either type of stability were to fail, then we would  reach a contradiction with the assumption, in the beginning of the argument, that the cascades have universal scaling. Such contradiction would imply  the absence of universal scaling. We have shown \cite{article:Gkioulekas:2008:1} that the inverse energy cascade is stable with respect to perturbations of the forcing statistics. The enstrophy cascade is only marginally stable, with the stability contingent on the smallness of the downscale energy flux. 

Third, from the fusion rules hypothesis and locality (a consequence of the fusion rule hypothesis) we have the argument of the present paper which leads to establishing the location of the dissipation scales, and, consequently, the existence of anomalous sinks of enstrophy and energy at small and large scales correspondingly. Our argument, more generally, establishes the relation between the energy dissipation rate $\gee_{ir}$ and the enstrophy dissipation rate $\gn_{uv}$ and the corresponding friction parameters $\gb$ and $\nu$ in terms  of the scaling exponents $\gz_2$ and $\xi_{2,1}$. As a matter of fact, we have seen that the scope of the argument is broader: we have derived the shape of the boundary separating the inertial range from the dissipation range in the $(r, R)$ plane representation for every generalized structure function $F_n$  for any $n\geq 2$.

As it stands now, the above argument has two loose ends. First, we have not worked out the problem of stability with respect to the sweeping interactions. As we have pointed out previously \cite{article:Gkioulekas:2007}, this problem, which was first recognized long ago by Kraichnan \cite{article:Kraichnan:1964}, remains an unresolved issue that affects every theoretical approach to the problem of turbulence that has been proposed to date. Second, the first step of the argument, establishing the fusion rules hypothesis from the assumption of universality, has not been done in a completely satisfactory manner.

This last point requires further clarification. The currently used formulation  of the hypothesis of universality is that the fundamental symmetries of the inertial range, namely  incremental homogeneity, incremental isotropy, and self-similarity, defined in terms of the  ensemble-averaged generalized structure functions, will survive imposing a symmetry-breaking restriction on the ensemble averaging \cite{article:Gkioulekas:2008:1}. Thus, the conditional generalized structure functions $\Phi_n$, defined via a conditional ensemble average of products of velocity differences incorporating the restriction, will still have an inertial range where the same symmetries are satisfied, as long as the symmetry breaking restriction is introduced at a scale that lies between the forcing scale and the scale defined by the velocity differences that are being averaged. Now, there is  a choice: The restriction can be placed on the forcing field or on the velocity-difference field. If the restriction is placed on the velocity-difference field, then the  fusion rules hypothesis is easily derived from the universality hypothesis. This was noted previously by L'vov and Procaccia \cite{article:Procaccia:1996:1,article:Procaccia:1996:3} in the context of three-dimensional turbulence, and I used the same argument to predict the fusion rules for both cascades of two-dimensional turbulence \cite{article:Gkioulekas:2008:1}. 

Now, this is hardly a surprise. From a physical standpoint, the hypothesis of universality, formulated in terms of a velocity-difference constraint on the ensemble, means that it makes no difference to the symmetries at a certain scale, if that scale is forced directly by gaussian forcing or indirectly by the velocity field at a larger (or smaller, for the inverse energy cascade) scale. In other words, by adopting the fusion rules hypothesis we are already, in a certain vague sense, hypothesizing the existence of a local cascade. Thus, the  locality and stability arguments, given previously \cite{article:Gkioulekas:2008:1}, and the arguments of the present paper, essentially establish a required consistency between the existence of local cascades and the governing Navier-Stokes equations. This is useful, but not good enough.

To establish the fusion rules hypothesis in a more definitive manner, we should start  from a weaker universality hypothesis in which 
we merely require that the scaling laws should not be sensitive to forcing perturbations, and nothing more. This is, after all, what we really intend to say when formulating the universality hypothesis. Arguments of this nature have been used to establish the fusion rules hypothesis for the passive scalar problem \cite{article:Procaccia:2000:1} of Kraichnan \cite{article:Kraichnan:1994} and the case $p = 2$ has been previously proved for the energy cascade of three-dimensional turbulence \cite{article:Procaccia:1995:1,article:Procaccia:1995:2,article:Procaccia:1996}. 
This type of stronger  argument  is currently missing for the case of the cascades in two-dimensions. 

This is why  the mathematical argument, both in this paper and in the previous paper \cite{article:Gkioulekas:2008:1}, was carefully formulated to account for  the case where the fusion scaling exponents $\xi_{n,p}$ get to   deviate from the values predicted by the fusion rules hypothesis. For example, in the previous paper \cite{article:Gkioulekas:2008:1} I investigated the effect a violation of the fusion rules will have on the locality of nonlinear interactions in the inertial range, and derived the required inequalities on the fusion  scaling exponents $\xi_{n,p}$ for preserving locality.  In the present paper, I derived the general dependence of the downscale enstrophy dissipation rate $\gn_{uv}$  on the viscosity $\nu$ and the upscale energy dissipation rate  $\gee_{ir}$  on the viscosity $\gb$  in terms of the  scaling exponent $\gz_2$  and the fusion  scaling exponent $\xi_{2,1}$. For the case where $\xi_{2,1}$  satisfies the fusion rules hypothesis and $a_3=1$, $\gn_{uv}$ and $\gee_{ir}$ become independent of the viscosities $\nu$ and $\gb$, which implies the existence of anomalous sinks. However, it should be stressed that our analysis was deliberately done in a general way to cover  the case where the fusion rules hypothesis fails.

With all the aforementioned concerns notwithstanding, I believe that a big picture is beginning to emerge, which can sharpen our physical intuition towards a deeper understanding of the cascades of two-dimensional turbulence. Already, from the present argument, we have  gained some interesting insights about two-dimensional turbulence, which we shall  now  summarize.  First, we have shown that it is impossible to have an inertial range in an enstrophy cascade dissipated by molecular dissipation ($\gk=1$), if the enstrophy cascade has intermittency corrections, or if, for some other reason, it deviates from Kraichnan scaling into steeper slopes.  Consequently, it may be easier for the enstrophy cascade to converge under hyperdiffusion as opposed to molecular dissipation.  Second, we have shown that the logarithmic correction plays an essential role in ensuring that the inertial range of the enstrophy cascade is not entirely destabilized  by dissipation, when $\gk=1$. Third, for the case of the inverse energy cascade, we have shown that if there are intermittency corrections to the scaling exponents $\gz_n$ , then the scaling exponents must satisfy the inequality $ \gz_{n+1}-\gz_n < 2m+1 ,\;\forall n>2$ , with $m$ being the order of the hypodiffusion, in order for \emph{all} generalized structure functions $F_n$ to have an inertial range.  Last but not least, we have not only shown that the fusion rules imply the existence of anomalous sinks; we have also established that a possible small violation of the fusion rules can be compensated for by increasing the orders $\gk$ and $m$ of hyperdiffusion and hypodiffusion correspondingly. For the case of three-dimensional turbulence we have also pointed to a possible connection between the bottleneck problem \cite{article:Falkovich:1994:1,article:Matsumoto:2004} and a possible temporary regular fusion rules violation.

To conclude, it should be noted that our work has focused strictly on the mathematical problem of two-dimensional turbulence that is precisely two-dimensional. More realism would require the introduction of three-dimensional effects at small scales, and still unaddressed remains the problem of how these effects impact on the dynamics of two-dimensional turbulence. In connection  with this very broad question, we can make the following remarks. First, the reason why we have a sink at large scales is precisely to account for Ekman damping, which is a three-dimensional large-scale frictional effect, known to exist for the case of atmospheric turbulence \cite{book:Pedlosky:1979}. A similar frictional effect exists in soap-film experiments of two-dimensional turbulence \cite{article:Wu:2000}. Second, at small enough scales, one expects a change in the dynamic of the nonlinear term. Based on what we have learned about locality from my previous paper \cite{article:Gkioulekas:2008:1}, I expect that the dynamic of the two-dimensional inertial ranges will be shielded from the nonlinear small-scale three-dimensional effect. However, the three-dimensional effect could change the nature of the small-scale dissipation. Further speculating on this issue is beyond the scope of this paper, though we have done so in previous papers \cite{article:Tung:2005,article:Tung:2005:1,article:Tung:2006,article:Tung:2007}. Let us just emphasize again that, based on our results here, we have learned that it makes a significant difference whether the enstrophy cascade is dissipated by diffusion or hyperdiffusion. As an anonymous referee noted, ``the short-wave dissipation is commonly considered as a parameterization which has to be designed so as to dissipate the enstrophy flux. In contrast, in the 3D case we believe that the true dissipation operator is known''. The fact that we do not know the true dissipation operator in two-dimensional  turbulence, in light of our results, is thus revealed to be a more urgent problem than is commonly realized.

\begin{acknowledgements}
The author wishes to thank an anonymous referee for his constructive remarks.
\end{acknowledgements}

\appendix

\section{Dissipation scales for an enstrophy cascade with intermittency corrections}
\label{app:enstr-with-int}

In this appendix we give the detailed derivation of the dissipation scales $\ell_{uv}^{(n)}(R)$ and $\gl_{uv}^{(n)}$ for the case of the enstrophy cascade with (mostly hypothetical) intermittency corrections.  To find the function $\ell_{uv}^{(n)}$, we write $\ell_{uv}^{(n)}(R) = \ell_{uv}(R) (R/\ell_0)^{a_n}$, and proceed to evaluate $a_n$ using the matching condition 
\begin{equation}
\frac{F_2 (\ell_{uv}^{(n)}(R))F_n (R)}{F_2 (R)}\sim \frac{(\ell_{uv}^{(n)}(R))^{2\gk}F_{n+1}(R)}{\nu R}.
\end{equation}
The factor $F_2 (\ell_{uv}^{(n)}(R))$ can be evaluated with judicious use of the properties of the dissipation scale functions as follows
\begin{align}
F_2(\ell_{uv}^{(n)}(R)) &\sim F_2 (\ell_{uv}(R)) (R/\ell_0)^{a_n\gz_2} \\
&\sim \frac{[\ell_{uv}(R)]^{2\gk} F_3(R)}{\nu R}(R/\ell_0)^{a_n\gz_2} \\
&\sim \frac{[\ell_{uv}^{(n)}(R)]^{2\gk}F_3(R)}{\nu R}(R/\ell_0)^{a_n\gz_2 -2a_n \gk} . 
\end{align}
 We substitute this into the matching condition and obtain  
\begin{equation}
\frac{(\ell_{uv}^{(n)}(R))^{2\gk}F_{n+1}(R)}{\nu R} \sim \frac{[\ell_{uv}^{(n)}(R)]^{2\gk}F_3(R)}{\nu R}(R/\ell_0)^{a_n\gz_2 -2a_n \gk} \frac{F_n (R)}{F_2 (R)} .
\end{equation}
If we employ the general evaluation of $F_3$ and $F_n$ we note that the factors $(\ell_{uv}^{n}(R))^{2k}$, $\nu R$, and $\gn_{uv}^{1/3}r$ cancel completely.  We are left with the balance
\begin{equation}
(R/\ell_0)^{\gz_{n+1}-(n+1)}  \sim (R/\ell_0)^{\gz_n -n-(\gz_2-2)+a_n (\gz_2 -2\gk)},
\end{equation}
that implies the scaling exponent equations
\begin{equation}
\gz_{n+1}-(n+1) = \gz_n -n-(\gz_2-2)+a_n (\gz_2 -2\gk).
\end{equation}
Solving for $a_n$ we get
\begin{equation}
a_n =  \frac{\gz_{n+1}-\gz_n+\gz_2 -3}{\gz_2-2\gk}. 
\end{equation}
We may now go back and write the dissipation scale function $\ell_{uv}^{(n)}(R)$ as follows
\begin{equation}
\begin{split}
\frac{\ell_{uv}^{(n)}(R)}{\ell_0} 
&=
\frac{\ell_{uv}(R)}{\ell_0}\left( \frac{R}{\ell_0}\right)^{a_n} = \left( \frac{R}{\ell_0}\right)^{x_n} \left [ \frac{\cR^{(\gn)}_{uv}}{\cR^{(\gn)}_{n,uv}}  \right]^{1/(\gz_2-2\gk)},
\end{split}
\end{equation}
where the scaling exponents $x_n$ are given by 
\begin{equation}
\begin{split}
x_n &= \frac{2}{\gz_2 -2\gk}+ \frac{\gz_{n+1}-\gz_n+\gz_2 -3}{\gz_2-2\gk}\\
&=\frac{\gz_{n+1}-\gz_n+\gz_2 -1}{\gz_2-2\gk}. 
\end{split}
\end{equation}

The standard dissipation scales for $F_n$ are found by solving the equation $\ell_{uv}^{(n)}(\gl_{uv}^{(n)})= \gl_{uv}^{(n)}$, leading to
\begin{equation}
\left( \frac{\gl_{uv}^{(n)}}{\ell_0}\right)^{x_n} \left [ \frac{\cR^{(\gn)}_{uv}}{\cR^{(\gn)}_{n,uv}}  \right]^{1/(\gz_2-2\gk)} = \fracp{\gl_{uv}^{(n)}}{\ell_0},
\end{equation}
and thus,
\begin{equation}
\fracp{\gl_{uv}^{(n)}}{\ell_0} =\left [ \frac{\cR^{(\gn)}_{uv}}{\cR^{(\gn)}_{n,uv}}  \right]^{1/[(1-x_n)(\gz_2-2\gk)]}.
\end{equation}
A simple calculation gives:
\begin{align}
(1-x_n)&(\gz_2-2\gk) \\
&= (\gz_2-2\gk)\cdot  \frac{(\gz_2-2\gk)-(\gz_{n+1}-\gz_n+\gz_2 -1)}{\gz_2-2\gk}\\
&=1-2\gk-(\gz_{n+1}-\gz_n),
\end{align}
and it follows that the dissipation scale $\gl_{uv}^{(n)}$ is given by
\begin{equation}
\frac{\gl_{uv}^{(n)}}{\ell_0} = \left [\frac{\cR^{(\gn)}_{uv}}{\cR^{(\gn)}_{n,uv}}\right]^{1/[1-2\gk-(\gz_{n+1}-\gz_n)]}.
\end{equation}

\begin{widetext}
\section{Dissipation scales for an enstrophy cascade without intermittency corrections}
\label{app:enstr-without-int}

In this appendix we give the detailed derivation of the dissipation scales $\ell_{uv}^{(n)}(R)$ and $\gl_{uv}^{(n)}$ for the case of the enstrophy cascade without intermittency corrections for the special case where $\gk =1$. Then, we confirm the admissibility condition. 

To find the function $\ell_{uv}^{(n)}$, we start by substituting $F_n (R) \sim (\gn_{uv}^{1/3}R)^n [\ln (\ell_0/R)]^{a_n}$ to the matching condition
\begin{equation}
\frac{F_2 (\ell_{uv}^{(n)}(R))F_n (R)}{F_2 (R)}\sim \frac{(\ell_{uv}^{(n)}(R))^{2}F_{n+1}(R)}{\nu R},
\end{equation}
which gives
\begin{equation}
\frac{(\gn_{uv}^{1/3}\ell_{uv}^{(n)}(R))^2 [\ln (\ell_0/\ell_{uv}^{(n)}(R))]^{a_2} (\gn_{uv}^{1/3}R)^n [\ln (\ell_0/R)]^{a_n}}{(\gn_{uv}^{1/3}R)^2 [\ln (\ell_0/R)]^{a_2}} 
\sim
\frac{\ell_{uv}^{(n)}(R)^2 (\gn_{uv}^{1/3}R)^{n+1} [\ln (\ell_0/R)]^{a_{n+1}}}{\nu R}. 
\end{equation}
After a few immediate cancellations we get the condition
\begin{equation}
\frac{1}{R^2}
\left [\ln\fracp{\ell_0}{\ell_{uv}^{(n)}(R)} \right ]^{a_2}
\left [\ln\fracp{\ell_0}{R}\right]^{a_n-a_2} 
\sim 
\frac{\gn_{uv}^{1/3}R}{\nu R} 
\left [\ln\fracp{\ell_0}{R}\right]^{a_{n+1}},
\end{equation}
from which we find that
\begin{align}
\left [\ln\fracp{\ell_0}{\ell_{uv}^{(n)}(R)} \right ]^{a_2}
& \sim 
\frac{\gn_{uv}^{1/3}R^2}{\nu} 
\left [\ln\fracp{\ell_0}{R}\right]^{a_{n+1}-a_n+a_2} 
\sim 
\frac{\gn_{uv}^{1/3}\ell_0^2}{\nu} 
\fracp{R}{\ell_0}^2
\left [\ln\fracp{\ell_0}{R}\right]^{a_{n+1}-a_n+a_2} \\
& \sim 
\cR_{uv}^{\gn}
\fracp{R}{\ell_0}^2
\left [\ln\fracp{\ell_0}{R}\right]^{a_{n+1}-a_n+a_2}.
\end{align}
Solving for the dissipation scale $\ell_{uv}(R)$, and introducing a critical Reynolds number $\cR^{(\gn)}_{2,uv}$, to turn the proportionality relation into an equation, gives the main result that reads:
\begin{equation}
\frac{\ell_{uv}^{(n)}(R)}{\ell_0}=
\exp\left [ - \fracp{\cR^{(\gn)}_{uv}}{\cR^{(\gn)}_{2,uv}}^{1/a_2}
\fracp{R}{\ell_0}^{2/a_2}
\left [\ln\fracp{\ell_0}{R}\right]^{b_n}\right ],
\end{equation}
with $b_n$   given by 
\begin{equation}
b_n = \frac{a_{n+1}-a_n+a_2}{a_2}.
\end{equation}

The standard dissipation scale $\gl_{uv}^{(n)}$ is the solution of the equation $\ell_{uv}^{(n)} (\gl_{uv}^{(n)}) = \gl_{uv}^{(n)}$, which reads
\begin{equation}
\ln\fracp{\ell_0}{\gl_{uv}^{(n)}} = 
\fracp{\cR^{(\gn)}_{uv}}{\cR^{(\gn)}_{2,uv}}^{1/a_2}
\fracp{\gl_{uv}^{(n)}}{\ell_0}^{2/a_2}
\left [\ln\fracp{\ell_0}{\gl_{uv}^{(n)}}\right]^{b_n},
\end{equation}
and then simplifies to the following transcendental equation 
\begin{equation}
\fracp{\ell_0}{\gl_{uv}^{(n)}}^{2/a_2}
\left [\ln \fracp{\ell_0}{\gl_{uv}^{(n)}}\right]^{1-b_n} = \fracp{\cR^{(\gn)}_{uv}}{\cR^{(\gn)}_{2,uv}}^{1/a_2}.
\end{equation}

 We may now confirm that the dissipation scale function $\ell_{uv}^{(n)}(R)$ satisfies the admissibility condition $\ell_{uv}^{(n)} (a\gl_{uv}^{(n)}) < a\gl_{uv}^{(n)}$ for all $a\gl_{uv}^{(n)} \in (\gl_{uv}^{(n)}, \ell_0)$. First, we note that $\ell_{uv}^{(n)}(\ell_0)=\ell_0$ (requires that $b_n > 0$) and $\ell_{uv}^{(n)}(\gl_{uv}^{(n)})=\gl_{uv}^{(n)}$. This means that the curve $r=\ell_{uv}^{(n)}(R)$ intersects the line $r=R$ at the endpoints of the inertial range $R=\gl_{uv}^{(n)}$ and $R=\ell_0$. The admissibility condition requires that when $R$ moves from $\gl_{uv}^{(n)}$ to $\ell_0$ the curve $r=\ell_{uv}^{(n)}(R)$ must remain under the line  $r=R$ throughout the interval $R \in (\gl_{uv}^{(n)}, \ell_0)$.   To confirm this behaviour, consider the partial derivative of $\ell_{uv}^{(n)}(R)$ with respect to $R$:
\begin{align}
\pD{R} \frac{\ell_{uv}^{(n)}(R)}{\ell_0} 
&= \pD{R}  \exp\left [ 
- \fracp{\cR^{(\gn)}_{uv}}{\cR^{(\gn)}_{2,uv}}^{1/a_2}
\fracp{R}{\ell_0}^{2/a_2}
\left [\ln\fracp{\ell_0}{R}\right]^{b_n}\right ] \\
&= - \frac{\ell_{uv}^{(n)}(R)}{\ell_0} \fracp{\cR^{(\gn)}_{uv}}{\cR^{(\gn)}_{2,uv}}^{1/a_2} 
\left\{ \frac{2}{a_2} \fracp{R}{\ell_0}^{\frac{2}{a_2}-1}\frac{1}{\ell_0} \left [\ln\fracp{\ell_0}{R}\right]^{b_n} 
 +  \fracp{R}{\ell_0}^{\frac{2}{a_2}} b_n \left [\ln\fracp{\ell_0}{R}\right]^{b_n-1} \frac{1}{\ell_0/R}\frac{-\ell_0}{R^2} \right\} \\
&= - \frac{\ell_{uv}^{(n)}(R)}{\ell_0} \fracp{\cR^{(\gn)}_{uv}}{\cR^{(\gn)}_{2,uv}}^{1/a_2} \fracp{R}{\ell_0}^{\frac{2}{a_2}-1} \left[ \ln\fracp{\ell_0}{R}\right]^{b_n-1} \frac{1}{\ell_0}
\left\{ \frac{2}{a_2}  \ln\fracp{\ell_0}{R} - b_n  \right\}. 
\end{align}
\end{widetext}
All factors above are positive, including the logarithm $\ln (\ell_0/R)$ (since $R<\ell_0$),  except for the last factor that can be either positive or negative. To satisfy the admissibility condition, it is sufficient to require that $\ell_{uv}^{(n)}(R)$ be decreasing at $R=\gl_{uv}^{(n)}$. This occurs if and only if
\begin{equation}
\frac{2}{a_2}  \ln\fracp{\ell_0}{\gl_{uv}^{(n)}} - b_n > 0,
\end{equation}
which is satisfied if and only if  $\gl_{uv}^{(n)} < \ell_0 \exp (-b_n a_2/2).$ For $R>\gl_{uv}^{(n)}$, the curve of $\ell_{uv}^{(n)}(R)$ will continue to decrease for a while, and will then turn around to increase and catch up with the line $r=R$, since $\ell_{uv}^{(n)}(\ell_0) = \ell_0$. 

\section{Inequalities on logarithmic  scaling exponents}
\label{app:inequalities} 

In this appendix we show that the assumption that the logarithmic scaling exponents $a_2$ and $a_3$ satisfy  $a_2>0$ and $a_3>0$ implies that $b_n > 0$ for all $n \in \bbN$ with $n>1$. This is a consequence of the inequality $a_{n+1}-a_{n} \geq a_{n}- a_{n-1}$ which can be established by an argument that is analogous to the one used to establish similar  inequalities  between the regular scaling exponents $\gz_n$ \cite{book:Feller:1968,article:Frisch:1991,book:Frisch:1995,article:Gkioulekas:2008:1}.  

We begin by defining  $\gd\gz (R)$ as the absolute value of the  scalar vorticity difference:
\begin{equation}
\gd\gz (R) = |\gz (\bfx+R\bfe,t)-\gz (\bfx,t)|,
\end{equation}
where $\bfx\in\bbR^d$ is given and $\bfe$ is a unit vector. The proof is based on the following two assumptions:
(a) For a downscale cascade, in the limit $\ell_0\goto\infty$, $\gd\gz (R)$ scales as $\avg{[\gd\gz(R)]^n} \sim [\ln (\ell_0/R)]^{a_n}$. 
(b) For finite $\ell_0$ there is a range of scales where the above scaling law continues to hold as an intermediate asymptotic.

Let $p, q \in (1,+\infty)$ with $1/p+1/q=1$, and let $\phi, \psi$ be two random variables with $\phi > 0$ and $\psi > 0$. The \Holder inequality for ensemble averages states that $\avg{\phi\psi} \leq \avg{\phi^p}^{1/p} \avg{\psi^q}^{1/q}$. For $p=q=1/2$ it reduces to the Schwarz inequality: $\avg{\phi\psi}^2 \leq \avg{\phi^2}\avg{\psi^2}$. We begin by choosing $\phi = [\gd\gz (R)]^{(n-1)/2}$ and $\psi =  [\gd\gz (R)]^{(n+1)/2}$ and employing the Schwarz inequality. It follows that
\begin{align}
\avg{[\gd\gz (R)]^n}^2 &= \avg{\phi\psi}^2 \leq \avg{\phi^2}\avg{\psi^2} \\
&= \avg{[\gd\gz (R)]^{n-1}}\avg{[\gd\gz (R)]^{n+1}},
\end{align}
and therefore
\begin{equation}
\frac{\avg{[\gd\gz (R)]^n}^2}{\avg{[\gd\gz (R)]^{n-1}}\avg{[\gd\gz (R)]^{n+1}}} \sim [\ln (\ell_0/R)]^{2a_n-a_{n-1}-a_{n+1}} < 1.
\end{equation}
To satisfy this inequality under the limit $\ell_0\goto\infty$ we require $2a_n-a_{n-1}-a_{n+1} \leq 0$ which is equivalent to $a_{n+1} - a_n \geq a_n - a_{n-1}$.

Let us assume now that $a_2>0$ and $a_3>0$. It follows that
\begin{align}
b_n &= \frac{a_{n+1}-a_n+a_2}{a_2} \\
&\geq \frac{a_{3}-a_2+a_2}{a_2} = \frac{a_{3}}{a_2} > 0,
\end{align}
which concludes the proof.

\bibliography{references}
\bibliographystyle{lfunsrt}

\end{document}